\documentclass[letterpaper,onecolumn,11pt,accepted=2020-09-07]{quantumarticle} %
\pdfoutput=1
\pdfoutput=1
\usepackage[english]{babel}

\usepackage{amssymb,amsmath,amsthm,amsfonts}
\usepackage{mathtools}
\usepackage{enumitem}
\usepackage[numbers,comma,sort&compress]{natbib}
\usepackage{authblk}
\usepackage{graphicx}
\usepackage[font=small]{caption}
\usepackage[labelformat=simple]{subcaption}
\usepackage{float}
\usepackage{xcolor}

\providecommand{\ketbra}[1]{\lvert  #1\rangle \langle #1 \rvert}
\newcommand{\expec}[1]{\langle #1\rangle}
\providecommand{\ket}[1]{\rvert #1 \rangle}

\begin{document}

\title{Entanglement characterization using quantum designs}

\author[aff1,aff2,aff4]{Andreas Ketterer}
\orcid{0000-0002-9045-692X} 
\author[aff3,aff4]{Nikolai Wyderka}
\orcid{0000-0003-3002-9878}
\author[aff4]{Otfried G\"uhne}
\orcid{0000-0002-6033-0867}
\affiliation[aff1]{Physikalisches Institut, Albert-Ludwigs-Universit\"at Freiburg,  Hermann-Herder-Str.~3, 79104~Freiburg, Germany}
\affiliation[aff2]{EUCOR Centre for Quantum Science and Quantum Computing, Hermann-Herder-Str.~3, 79104~Freiburg, Germany}
\affiliation[aff3]{Institut f\"ur Theoretische Physik~III, Heinrich-Heine-Universit\"at D\"usseldorf, Universit\"atsstr.~1, 40225~D\"usseldorf, Germany}
\affiliation[aff4]{Naturwissenschaftlich-Technische Fakult\"at, Universit\"at Siegen, Walter-Flex-Str.~3, 57068 Siegen, Germany}
\date{}

\begin{abstract}
We present in detail a statistical approach for the reference-frame-independent detection 
and characterization of multipartite entanglement based on moments of randomly measured 
correlation functions. We start by discussing how the corresponding moments can be evaluated 
with designs, linking methods from group and entanglement theory. Then, we illustrate the 
strengths of the presented framework with a focus on the multipartite scenario. We discuss
a condition for characterizing genuine multipartite entanglement for three qubits, and we 
prove criteria that allow for a discrimination of $W$-type entanglement for an 
arbitrary number of qubits.
\end{abstract}


\maketitle


\section{Introduction}

The experimental detection of multipartite entanglement usually requires a number of 
appropriately chosen local quantum measurements which are aligned with respect to a 
previously shared common reference frame~\cite{OtfriedReview,ReviewRefFrames}. The latter, 
however, can be a challenging prerequisite for photonic free-space quantum communication 
over distances of several hundreds of kilometers~\cite{FreeSpaceQC,FreeSpaceComm}, which 
is currently in the process of being extended to space involving satellites orbiting the earth~\cite{SatelliteQC1,SatelliteQC2,SatelliteQC3,SatelliteQC4}. Here, due to the motion, 
distance and number of involved satellites, the issue of sharing classical reference frames 
becomes particularly challenging, making the development of alternative detection strategies desirable.

In recent years, there has been a number of proposals of experimental protocols that avoid the need 
of sharing classical reference frames. One possibility is to encode logical qubits into rotational 
invariant subspaces of combined degrees of freedom of photons, i.e. their polarization and transverse 
degrees of freedom \cite{LeandroRotInv,SteveAlignmentFreeComm}. While the latter procedure provides 
one with a complete experimental toolbox for alignment free quantum communication, one can also find 
experimentally less demanding strategies that allow for the reference frame independent certification 
of entanglement.  
For instance, one can use entanglement criteria that are invariant under local unitary (LU) 
transformations, commonly termed as reference-frame-independent~\cite{BriegelLUinv,JulioCorrMat1,JulioCorrMat2,JulioMarkus,ZukowskiRefFrame1,ZukowskiRefFrame2,LawsonRefFrame,KlocklHuber}. 
This type of entanglement criteria requires that the experimenters are capable of measuring a fixed set of 
local observables, but  completely avoids the need of aligning measurements among different parties.  

\begin{figure}[t!]
\begin{center}
\includegraphics[width=0.8\textwidth]{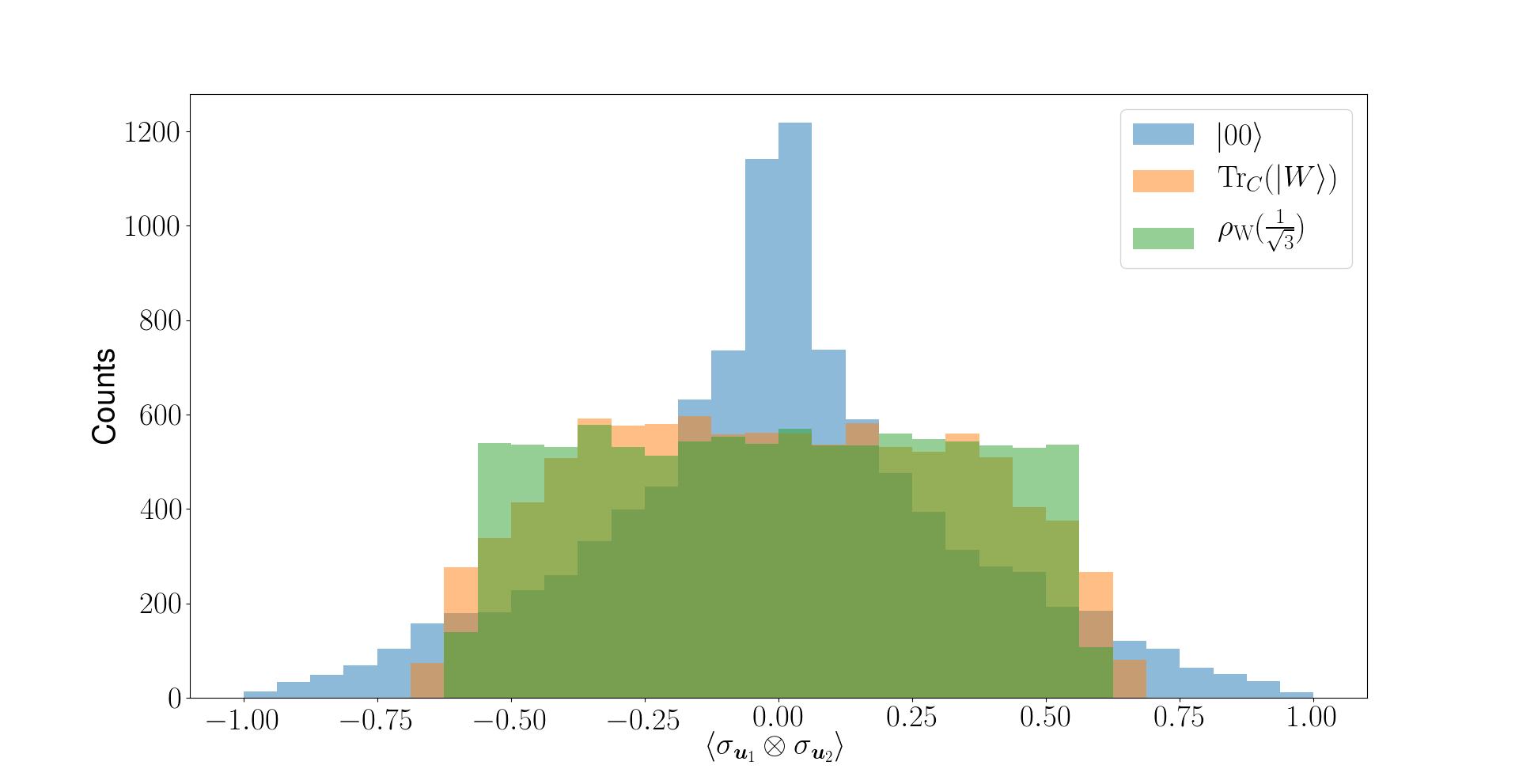}%
\end{center}
\caption{Example for entanglement detection using the statistics of 
random measurements. For a given two-qubit state, one can measure
the expectation value 
$E({\boldsymbol u}_1,{\boldsymbol u}_2)=\langle \sigma_{{\boldsymbol u}_1}\otimes \sigma_{\boldsymbol u_2} \rangle$ 
in randomly chosen local directions ${\boldsymbol u}_1$ and ${\boldsymbol u_2}$. The 
figure shows a Histogram of the observed counts for $10^4$ measured values of $E({\boldsymbol u}_1,{\boldsymbol u}_2)$ with uniformly sampled directions ${\boldsymbol u}_1$ and ${\boldsymbol u}_2$. Colors indicate three different two qubit states: product state (blue), 
the two-body marginal state of the tripartite $W$-state (orange) and a specific 
Werner state (green). The distributions share the same mean value and, in fact, 
also the same variance (as one has $\mathcal{R}^{(2)}=1/9$ in Eq.~(\ref{eq:RandomMoments})),
but the forth moment $\mathcal{R}^{(4)}$ differs and allows to detect the entanglement
of the Werner state and the marginal state.}
\label{figure_1}
\end{figure}

In several recent works it has been shown how to go beyond such procedures by relaxing also 
the assumption of being able to measure a fixed set of local observables and instead allow 
only for local measurements with settings drawn uniformly at random~\cite{tran1,tran2,
MeMoments,MichaelBachelor,ZollerFirst,ZollerScience,ElbenPRA,ElbenPRL, MeineckeExperimentRandom}. 
The common idea of these approaches is to measure a certain correlation function,
and average the result over random local unitaries applied to the state. Clearly, 
in this scenario one has to resort to statistical tools based on the moments of the
resulting probability distribution of correlations in order to infer the nonlocal 
properties of the underlying quantum states (see also Fig.~\ref{figure_1}). Furthermore, apart from its reference-frame-independent nature, such protocols are advantageous for the characterization of large multipartite systems where a complete reconstruction of the underlying quantum state becomes practically impossible due to the required measurement resources. 

So far, however, most of the aforementioned 
approaches have been focused on the lowest statistical moments
only~\cite{tran1,tran2,ZollerFirst,ZollerScience,ElbenPRA,ElbenPRL}, and applied them to the detection 
of either the absence of full-separability~\cite{tran1,tran2}, or bi-separability with 
respect to a predetermined bipartition~\cite{ZollerFirst,ZollerScience,ElbenPRA,ElbenPRL}. 
Though the latter approaches have proven useful for experimental implementations 
in trapped ion experiments \cite{ZollerScience,ElbenPRL}, it is important to note
that the knowledge of all moments corresponding to different sectors, i.e., subsets 
of the involved parties, gives more insight into the entanglement properties of 
the underlying state. For the case of second moments, this follows from the relation 
between the respective moments and so-called sector lengths which have been 
independently under investigation in the context of entanglement detection~\cite{NikolaiConstraintsCorr,NikolaiSectorLengths,JensR2max}. Similar 
insights have been reached in Ref.~\cite{MeineckeExperimentRandom} where the detection 
and characterization of multipartite entanglement based on second moments 
was studied experimentally with entangled photons.

Moreover, it has recently been demonstrated that a combination of statistical moments 
beyond the second order can lead to further improvements in terms of entanglement 
detection~\cite{MeMoments}. In particular, it was shown that novel reference-frame-independent criteria for the detection and also characterization of multipartite 
entanglement can be derived by expressing the respective moments in terms of 
spherical designs, i.e., pseudo-random processes allowing to mimic uniform 
averages over the sphere. This insight also has experimental consequences because the evaluation of the respective moments in terms of spherical designs is exact and, for multipartite systems involving a small or intermediate sized number of parties, does not exploit too many measurement resources. Lastly, it was also shown that a quantification of bipartite entanglement in terms of the respective moments is possible as well~\cite{MichaelBachelor}.

In the present contribution  we will further investigate the potential of this 
framework. After recalling for completeness some of the results from Ref.~\cite{MeMoments},
we first discuss the characterization of multipartite entanglement classes based 
on the first two non-vanishing moments in small multipartite systems consisting 
of three and four qubits. Finally, we prove two novel criteria enabling 
the discrimination of $W$-type entangled mixed states for an arbitrary number of parties.

The paper is organized as follows. In Sec.~\ref{sec:TheoFrame} we introduce the 
necessary theoretical tools from Ref.~\cite{MeMoments}, i.e., the moments of 
random correlation functions and the concept of unitary and spherical designs (see Sec.~\ref{sec:RandomMoments} and \ref{sec:QuDesigns}, respectively). These
are then used in Sec.~\ref{sec:MomDes} to evaluate moments of random correlation 
functions of different order. Section~\ref{sec:ApplQubitSys} then focuses on 
the application of the introduced framework to few-qubit system discussing 
the characterization of multipartite entanglement based on the first two 
non-vanishing moments. Lastly, the criterion enabling the discrimination 
of $W$-type entanglement in multi-qubit systems is introduced in 
Sec.~\ref{sec:DiscWclass}. Finally, we conclude in  Sec.~\ref{sec:Conclusion} 
and give a short outlook.

 
\section{Theoretical framework}\label{sec:TheoFrame}

\subsection{Moments of random correlations}\label{sec:RandomMoments}
To set the stage we consider a system of $N$ $d$-dimensional quantum systems 
(qudits) prepared in the initial state $\rho$. Subsequently, each of the 
qudits is measured in a randomly drawn basis
\begin{align}
\left\{\left(\ket{u_n^{(0)}}:= U_n\ket{0_n},\ket{u_n^{(1)}}:= U_n\ket{1_n},\ldots,\ket{u_n^{(d-1)}}:= U_n\ket{(d-1)_n}\right)\right\}_{n=1,\ldots,N},
\end{align}
 each specified by a random unitary transformation $U_n$ picked uniformly from the unitary group $\mathcal U(d)$, i.e., according to the Haar measure. One round of such random measurements yields the corresponding correlation function $\expec{ U_1\mathcal O U_1^\dagger\otimes \ldots \otimes U_N\mathcal O U_N^\dagger}$, where $\mathcal O$ describes an arbitrary qudit observable diagonal in the computational basis $\{\ket{0_n},\ldots,\ket{(d-1)_n}\}$, and $\expec{...}$ denotes the expectation value with respect to the quantum state $\rho$. We note that in general the choice of $\mathcal O$ is relevant if one considers systems of local dimensions $d$. However, it turns out that in the case $d=2$,  which is the main focus of Secs.~\ref{sec:ApplQubitSys} and \ref{sec:DiscWclass}, any local qubit observable will suffice\footnote{This is a direct consequence of the isomorphism between SU$(2)/\mathbb Z_2$ and SO$(3)$.}. 
 
Further on, as the unitary transformations $U_n$  are chosen randomly, a single set of random measurement settings will not give much insight  into  the nonlocal properties of the initial state $\rho$. In order to achieve this we have to perform several rounds of random measurements and seek a statistical treatment in terms of the moments of the randomly measured correlation functions.  As we have assumed that each correlation function is characterized by a set of Haar random unitaries $\{U_n\}_{n=1,\ldots,N}$, we can define these moments as follows~\cite{MeMoments}:
\begin{align}
\mathcal R^{(t)}&= \int_{\mathcal U(d)} d\eta(U_1) \ldots \int_{\mathcal U(d)} d\eta(U_N)
\expec{ U_1\mathcal O U_1^\dagger\otimes \ldots \otimes U_N\mathcal O U_N^\dagger}^t,
\label{eq:RandomMomentsQudits}
\end{align}
where $t$ is a positive integer,  and $\eta$ the Haar measure on the unitary group $\mathcal U(d)$. It is important to note that one can gain more information about $\rho$ by considering also moments of smaller qubit sectors, i.e., the respective reduced states, as has been investigated for $t=2$ in Refs.~\cite{NikolaiConstraintsCorr,ZollerScience,ElbenPRA,ElbenPRL,NikolaiSectorLengths,JensR2max,MeineckeExperimentRandom}. However, in the remainder of this manuscript we will focus on the characterization of multipartite entanglement based on full $N$-qubit moments~(\ref{eq:RandomMomentsQudits}).

Further on, we note that the random observables $U\mathcal O U^\dagger$, with $U\in \mathcal U(d)$, can be parametrized by $d(d-1)$ angles. This follows directly from the fact that any $U\in \mathcal U(d)$ can be decomposed as $U=e^{i \varphi}Z_1XZ_2$, where $\varphi$ is a global phase, $Z_1$ and $Z_2$ are diagonal unitary matrices with $[Z_1]_{11}=[Z_2]_{11}=1$ and $X$ is a unitary matrix with $\sum_i [X]_{ij}=\sum_i [X]_{ji}=1$, for all $j$ \cite{DecompositionU}.  We thus have that $U\mathcal O U^\dagger=Z_1X\mathcal O X^\dagger Z_1^\dagger$ and, since the matrices $A$ are isomorphic to $\mathcal U(d-1)$, a simple count of parameters leads to $d(d-1)$. Hence, in the case of two-level systems (qubits) we find that the local measurement settings are characterized by only two angles corresponding to the spherical coordinates fixing a direction on the Bloch sphere $S^2$. In other words, we can  associate to each random basis $(\ket{u_n^{(0)}}:= U_n\ket{0_n},\ket{u_n^{(1)}}:= U_n\ket{1_n})$, a direction 
$\boldsymbol u_n\in S^2$, defined by the components $[\boldsymbol u_n]_i=\mathrm{tr}[\sigma_{\boldsymbol u_n}\sigma_i]/2$, where $\sigma_i$, with $i=x,y,z$, denote the usual Pauli matrices and $\sigma_{\boldsymbol u_n}= U_n \sigma_z U_n^\dagger$. Equation~(\ref{eq:RandomMomentsQudits}) for the $t$-th moment thus becomes 
\begin{align}
\mathcal R^{(t)}&= \int_{\mathcal U(2)} d\eta(U_1) \ldots \int_{\mathcal U(2)} d\eta(U_N)
\expec{ U_1\sigma_{z}U_1^\dagger\otimes \ldots \otimes U_N\sigma_{z}U_N^\dagger}^t\label{eq:RandomMoments1}\\
&= \frac{1}{(4\pi)^N} \int_{S^{2}} d\boldsymbol u_1\ldots \int_{S^{2}} d\boldsymbol u_N E(\boldsymbol u_1,\ldots,\boldsymbol u_N)^t,
\label{eq:RandomMoments}
\end{align}
where we defined $E(\boldsymbol u_1,\ldots,\boldsymbol u_N):=\expec{\sigma_{\boldsymbol u_{1}}\otimes \ldots \otimes \sigma_{\boldsymbol u_{N}}}$, and $d\boldsymbol u_i=\sin{\theta_i} d\theta_i d\phi_i$ denotes the uniform measure on the Bloch sphere $S^{2}$.
In particular, it is easy to see from Eq.~(\ref{eq:RandomMoments})  that all odd moments are zero due to the symmetry of the correlation functions $E(\boldsymbol u_1,\ldots,\boldsymbol u_N)$ with respect to a reflection on the Bloch sphere: $E(\boldsymbol u_1,\ldots,-\boldsymbol u_i,\ldots, \boldsymbol u_N)=-E(\boldsymbol u_1,\ldots,\boldsymbol u_N)$. 

In the following, we show that the moments $\mathcal R^{(t)}$ can be calculated using unitary $t$-designs (or spherical $t$-designs for $d=2$), rather than averaging over the whole unitary group.

\subsection{Designs}\label{sec:QuDesigns}

Generally speaking quantum designs are pseudo-random processes that allow to mimic uniform averages over some group if one is only interested in moments up to some finite degree. Depending on the specific choice of the group such processes are either referred to as unitary or spherical designs. In the following, we will give short introduction to the two of them.
\subsubsection{Unitary designs}
Let us denote by $\mathrm{Hom}(r,s)$ the set of all homogeneous polynomials $P_{r,s}(U)$, with support on the space of unitary matrices $\mathcal U(d)$, that is of degree at most $r$ and $s$, respectively, in each of the matrix elements of $U$ and their complex conjugates. For example, the polynomial $P_{r=1,s=2}(U)=U^\dagger V U V U^\dagger$ is of degree $r=1$ and $s=2$ in the entries of the matrix $U$. 
With this we arrive at the following definition~\cite{Dankert}:
A {unitary $t$-design} is a set of unitary matrices $\{U_k|k=1,\ldots,K^{(t)}\}\subset \mathcal U(d)$, with cardinality $K^{(t)}$, such that 
\begin{align}
\frac{1}{K^{(t)}} \sum_{k=1}^{K^{(t)}} P_{t',t'}(U_k) =\int_{\mathcal U(d)} P_{t',t'}(U) d\eta(U),
\label{eq:UnitaryDesDef}
\end{align} 
for all homogeneous polynomials $P_{t',t'}\in \mathrm{Hom}(t',t')$, with $t'\leq t$, and where $\eta(U)$ denotes the normalized Haar measure on $\mathcal U(d)$.

We note that, while the existence of unitary designs has been proven \cite{existence}, no universal strategy for their construction in case of an arbitrarily given $t$ exists. This fact led to the study of approximate unitary designs for which the property (\ref{eq:UnitaryDesDef}) is accordingly relaxed \cite{ApproxDesign1,ApproxDesign2,ApproxDesign3}. However, in the remainder of this manuscript we will restrict ourselves to particular cases in which exact design are known. A prominent example of an exact unitary $t$-design is given by the multi-qubit Clifford group consisting of all unitary matrices mapping the multi-qubit Pauli group onto itself.  The latter has been shown to constitute a unitary $3$-design~\cite{Cliff3Design}, however, it fails to be a unitary $4$-design~\cite{CliffNo4Design}. In the case of a single qubit the Clifford group has $24$ elements which can be generated from the Hadamard gate $ H$ and the phase gate $ S=e^{i \frac{\pi}{4} \sigma_z}$. 
Furthermore, in Ref.~\cite{Gross5Design} the existence of a qubit $5$-design of one qubit was noted. The latter is given by the unitary representation of the special linear group $SL(2,\mathbb F_5)$ of invertible $2\times 2$ matrices over the finite field $\mathbb F_5$ with five elements. In App.~\ref{app:Unitary5Design} we shortly outline how to generate this design.

\subsubsection{Spherical designs}
\begin{figure}[t!]
\begin{center}
\includegraphics[width=0.6\textwidth]{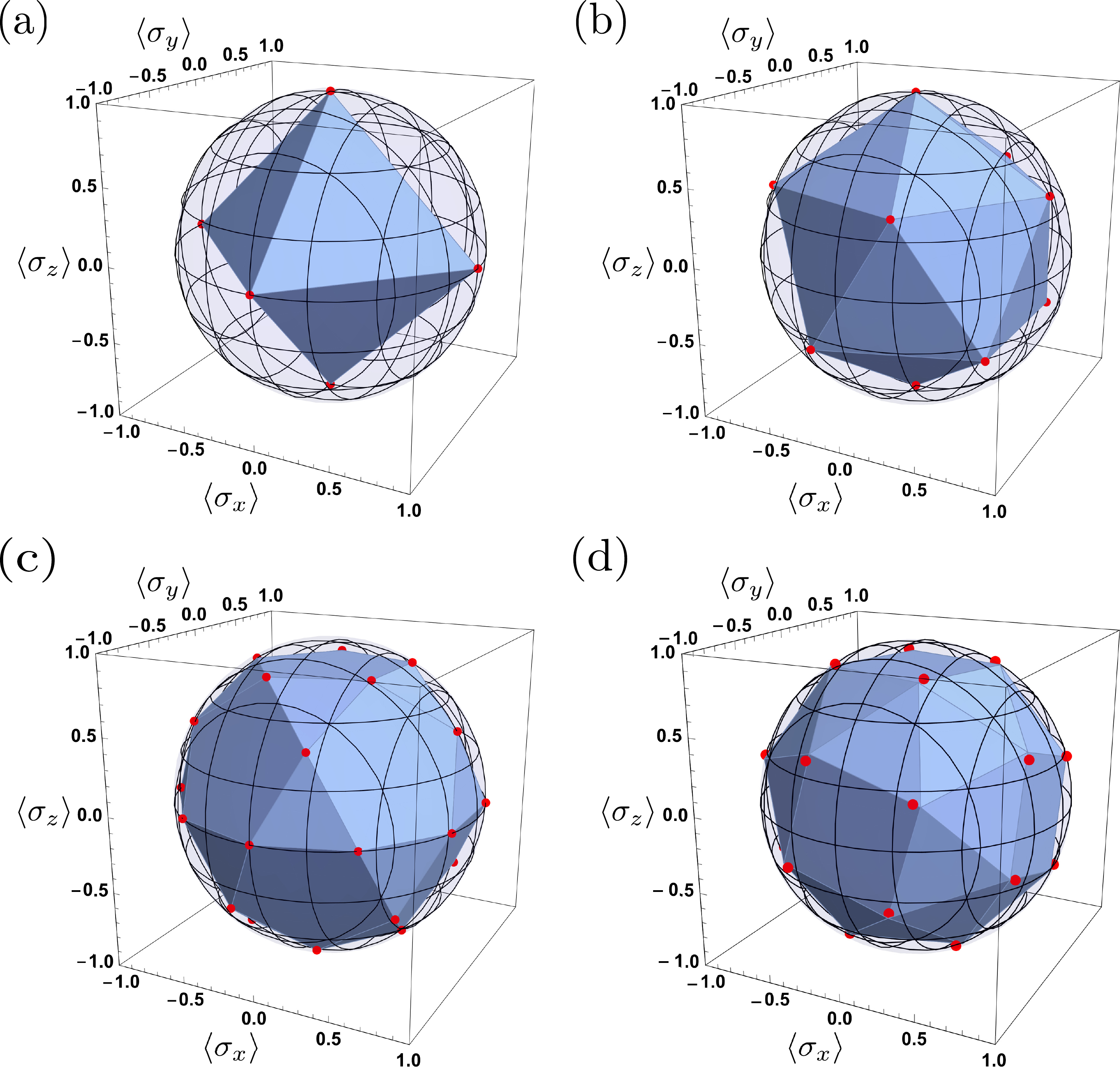}%
\end{center}
\caption{Plot of the Bloch vectors of various spherical designs, for $d=2$, and their corresponding polyhedra. The components of the Bloch vector are denoted as $\expec{\sigma_i}$, with the Pauli matrices $\sigma_x$, $\sigma_y$, and $\sigma_z$. (a)  $3$-design with $K_\text{octa}=6$ vertices forming an octahedron. (b) $5$-design with $K_\text{icosa}=12$ vertices forming an icosahedron. (c) $5$-design with $K_\text{icosi}=30$ vertices forming an icosidodecahedron. (d) $7$-design with $K_{7-\text{design}}=24$ vertices forming a deformed snub cube (red points). For comparison, a regular snub cube only forms a $3$-design (blue polyhedron).}
\label{figure_2}
\end{figure}
In Sec.~\ref{sec:RandomMoments} we saw that in the case of systems consisting of qubits the evaluation of the moments $\mathcal R^{(t)}$ boils down to local integrations over the spheres $S^2$. In this case, instead of using a unitary $t$-design to evaluate the respective moments, we can resort to the concept of spherical designs. In general, a spherical $t$-design in dimension three consist of a finite set of points $\{\boldsymbol u_{i}|i=1,\ldots,L^{(t)}\}\subset S^2$ fulfilling the property 
\begin{align}
\frac{1}{L^{(t)}} \sum_{k=1}^{L^{(t)}} P_{t'}(\boldsymbol u_k) = \frac{1}{4\pi}\int_{S^{2}}d\boldsymbol u \ P_{t'}(\boldsymbol u),
\label{eq:tDesignDef}
\end{align} 
for all homogeneous polynomials $P_{t'}:S^2\rightarrow \mathbb R$, with $t'\leq t$. It thus suffices to resort to spherical $t$-designs as long as one is interested in calculating averages of polynomials of degree at most $t$ over the Bloch sphere $S^2$. 

One way to generate spherical designs is to extract them from unitary designs. For instance, by applying the elements of the single-qubit Clifford group to one of the Pauli matrices, e.g. $\sigma_z$, we are left with the following set of inequivalent operators $\{\pm \sigma_x,\pm\sigma_y,\pm\sigma_z\}$. The latter correspond to the following set of unit vectors $\{\pm\boldsymbol e_i|i=x,y,z\}$ which form a spherical $3$-design (see Fig.~\ref{figure_2}(a)). Similarly, one can generate a spherical $5$-design from the $60$ element unitary $5$-design $SL(2,\mathbb F_5)$. To do so, we calculate again all inequivalent directions on the Bloch sphere originating from the operators $\sigma_{{\boldsymbol u}^{(k)}}=U^{(k)}\sigma_z{U^{(k)}}^\dagger$, for all elements $U^{(k)}$ of the unitary $5$-design. The result is a set of 30 vertices on the Bloch sphere forming an icosidodecahedron (see Fig.~\ref{figure_2}(c)). 

In general, however, spherical designs are easier to find than unitary designs because it is easier to search them directly by checking the relation~(\ref{eq:tDesignDef}) for sets of vertices on the sphere $S^2$. 
Such a search was carried out in Ref.~\cite{ExamplesSphericalDesigns} where a number of spherical designs on the $2$-sphere $S^2$ for $t$'s up to 20 and consisting of up to 100 elements were found. In Fig.~\ref{figure_2}(b) and (d) we present two more examples of such spherical designs. The $5$-designs presented in Fig.~\ref{figure_2}(b) corresponds to an icosahedron with $K_\text{icosa}=12$ vertices. Furthermore, the $7$-design with $K_{7-\text{design}}=24$ vertices, shown in Fig.~\ref{figure_2}(d), represents a deformed snub cube, i.e., a regular snub cube with slightly smaller square faces and slightly larger triangular faces~\cite{ExamplesSphericalDesigns}. Interestingly, the icosidodecahedron, presented in Fig.~\ref{figure_2}(c), and the regular snub cube, presented in blue in Fig.~\ref{figure_2}(d), constitute only spherical $5$- and $3$-designs, respectively.

Lastly, we emphasize that for systems of larger local dimensions it is difficult to resort to spherical designs instead of unitary ones for the evaluation of the moments $\mathcal R^{(t)}$. This is due to the fact that the parametrization of the space of observables of the form $U\mathcal O U^\dagger$ (see Sec.~\ref{sec:RandomMoments}), is not in one to one correspondence with points on a generalized Bloch sphere $S^{d^2-1}$. It rather forms a submanifold of $S^{d^2-1}$ characterized by $d(d-1)$ parameters. Hence, generalized spherical designs on $S^{d^2-1}$ will not be useful in this case. 

\section{Random moments from designs}\label{sec:MomDes}

We first note that the $t$-th power of the correlation function $\expec{ U_1\mathcal O U_1^\dagger\otimes \ldots \otimes U_N\mathcal O U_N^\dagger}$ is a polynomial of degree $t$ in the entries of the local random unitary matrices $U_n\in \mathcal U(d)$, and their complex conjugates. Hence, in order to evaluate the moments $\mathcal R^{(t')}$, with $t'\leq t$, it suffices to average $\expec{ U_1\mathcal O U_1^\dagger\otimes \ldots \otimes U_N\mathcal O U_N^\dagger}^{t'}$ locally over a respective unitary $t$-design instead over the whole unitary group $\mathcal U(d)$. 
This leads to the result
\begin{align}
\mathcal R^{(t')}&=\frac{1}{(K^{(t)})^N}\sum_{k_1,\ldots,k_N=1}^{K^{(t)}}\expec{ U_{k_1}\mathcal O U_{k_1}^\dagger\otimes \ldots \otimes U_{k_N}\mathcal O U_{k_N}^\dagger}^{t'},
\label{eq:MomentUnitaryDesign}
\end{align}  
for $t'\leq t$, and with a unitary $t$-design $\{U_n\}_{k=1}^{K^{(t)}}$.  Equation~(\ref{eq:MomentUnitaryDesign}) is a general formula that allows one to calculate the moments (\ref{eq:RandomMoments}) up to a certain $t$ provided a unitary $t$-design can be found. In the remainder of this paper we will focus on the qubit case ($d=2$) for which Eq.~(\ref{eq:MomentUnitaryDesign}) takes a particularly simple form. In this case Eq.~(\ref{eq:MomentUnitaryDesign}) becomes 
\begin{align}
\mathcal R^{(t)}
&=\frac{1}{(L^{(t)})^N} \sum_{k_1,\ldots,k_N=1}^{L^{(t)}} \expec{\sigma_{\boldsymbol u_{k_1}}\otimes \ldots \otimes \sigma_{\boldsymbol u_{k_N}}}^t,
\label{eq:MomentSphericalDesign}
\end{align}  
where $L^{(t)}\leq K^{(t)}$ denotes the count of the remaining non-equivalent measurement directions $\{\boldsymbol u_i\}_{i=1,\ldots,L^{(t)}}$, after dropping those for which $j$'s exist  with $\sigma_{\boldsymbol u_{k_i}}=\sigma_{\boldsymbol u_{k_j}}$.
Hence, in order to calculate the moments $\mathcal R^{(t)}$ for systems of qubits it is enough to average  $\left[E(\boldsymbol u_1,\ldots,\boldsymbol u_N)\right]^t$ over a finite number $L^{(t)}$ of nonequivalent Bloch sphere directions $\{\boldsymbol u_{i}|i=1,\ldots,L^{(t)}\}\subset S^2$ which themselves form a spherical $t$-design. 

We explicitly evaluate the moments (\ref{eq:MomentSphericalDesign}) using the $6$ element spherical $3$-design and $12$ element spherical $5$-design, presented in Sec.~\ref{sec:QuDesigns}, yielding:
\begin{align}
\mathcal R^{(2)}&=\frac{1}{3^N} \sum_{i_1,\ldots,i_N=x,y,z} E(\boldsymbol e_{i_1}, \ldots  \boldsymbol e_{i_N})^2,\label{eq:RandomMoment2}  \\
\mathcal R^{(4)}&=\frac{1}{6^N} \sum_{i_1,\ldots,i_N=1}^{6} E(\boldsymbol v_{i_1}, \ldots  \boldsymbol v_{i_N})^4,
\label{eq:RandomMoment4}
\end{align}
with the upper summation bounds given by $L^{(3)}/2=3$ and $L^{(5)}/2=6$, respectively.  
Note that in Eqs.~(\ref{eq:RandomMoment2}) and (\ref{eq:RandomMoment4}) the number of summands is $L^{(t)}/2$ because for even $t$ one can drop the respective anti-parallel settings $-\boldsymbol e_i$ and $-\boldsymbol v_i$. Also note that we always use those spherical designs with the smallest number of elements, in order to minimize the number of involved measurement settings.

In the remainder of the manuscript we will mainly focus on entanglement criteria involving the first two non-vanishing moments, i.e., $\mathcal R^{(2)}$ and $\mathcal R^{(4)}$. However, we emphasize that is generally possible to evaluate also higher order moments using higher order designs. For instance, with the deformed snub cube spherical $7$-design $\{\boldsymbol w_i|i=1,\ldots,L^{(7)}=24\}$ (see Fig.~\ref{figure_2}(d)) we obtain the following formula for  the sixth moment 
\begin{align}
\mathcal R^{(6)}=\frac{1}{24^N} \sum_{i_1,\ldots,i_N=1}^{24} E(\boldsymbol w_{i_1}, \ldots  \boldsymbol w_{i_N})^6.
\label{eq:sixthmoment}
\end{align}
Note that the number of summands in Eq.~(\ref{eq:sixthmoment}) cannot be reduced to $L^{(7)}/2$, as in Eqs.~(\ref{eq:RandomMoment2}) and (\ref{eq:RandomMoment4}), because the deformed snub cube is not point symmetric.

\section{Applications to qubit systems}
\label{sec:ApplQubitSys}

\subsection{Two qubits}\label{sec:2QubitStuff}
A general two-qubit density matrix can be expressed in the following form
\begin{align}
\rho=\frac{1}{4}\big[\mathbbm 1_4  + (\boldsymbol a\cdot \boldsymbol\sigma )\otimes \mathbbm 1_2+\mathbbm 1_2 \otimes (\boldsymbol b\cdot \boldsymbol\sigma)+\sum_{i,j=x,y,z} c_{i,j} \sigma_i \otimes \sigma_j \big],
\label{eq:2QuFano}
\end{align}
where $\boldsymbol\sigma=(\sigma_x,\sigma_y,\sigma_z)^{\top}$ denotes the Pauli spin operator, $\boldsymbol a$ and $\boldsymbol b$ the local Bloch vectors of the reduced subsystems of each individual qubit, and $C=\{c_{i,j}\}_{i,j=x,y,z}=\{\text{tr}[\rho \sigma_i\otimes\sigma_j]\}_{i,j=x,y,z}$ the correlation matrix.  
To start, we focus first on so-called Bell diagonal states, i.e. states that are diagonal in the Bell basis and whose density matrix can be expressed in the following way  $\rho_{\text{BD}}=\frac{1}{4}\big[\mathbbm 1_4  +\sum_{j=x,y,z} c_j \sigma_j \otimes \sigma_j \big]$, with real parameters $c_j$, such that $0\leq |c_j|\leq 1$, and the corresponding eigenvalues $\lambda_{1,2}=(1\mp c_x \mp c_y-c_z)/4$ and $\lambda_{3,4}=(1\pm c_x \mp c_y+c_z)/4$. Bell diagonal states are separable if and only if $|c_x|+|c_y|+|c_z|\leq 1$~\cite{BellDiagStates}.

\begin{figure}[t]
\begin{center}
\includegraphics[width=\textwidth]{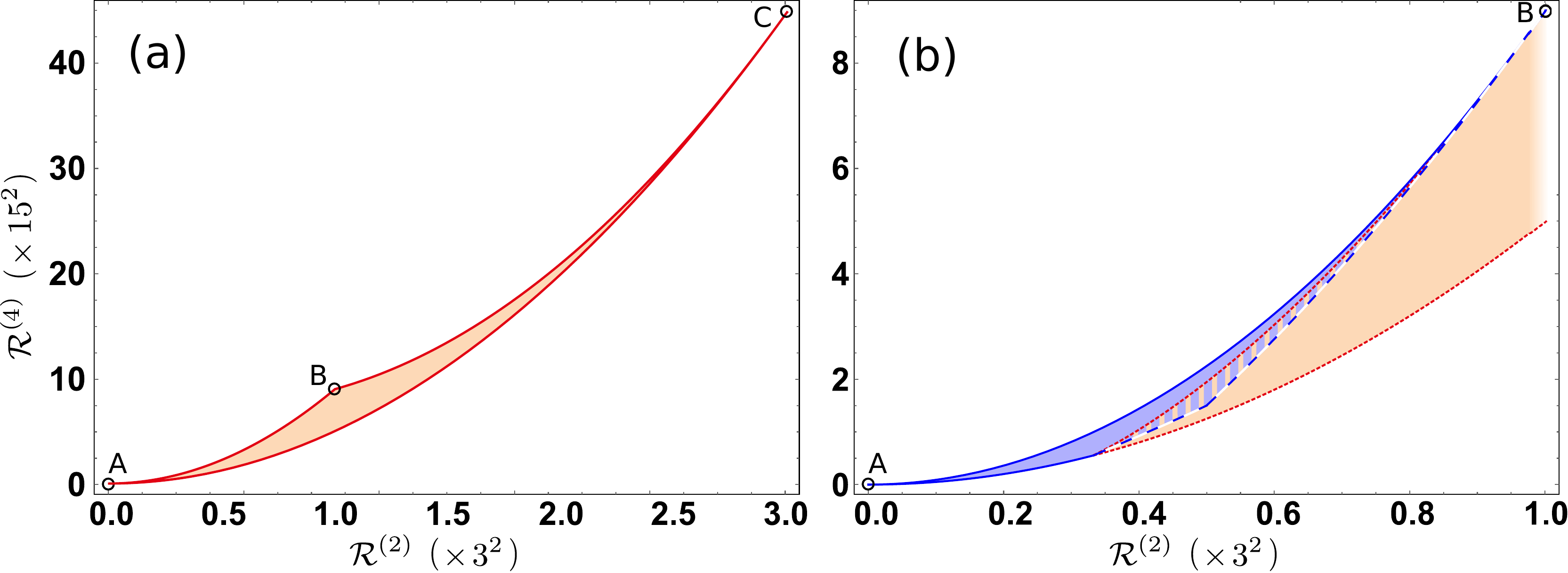}%
\end{center}
\caption{(a) Representation of the set of Bell diagonal states in the space spanned by the moments $\mathcal R^{(2)}$ and $\mathcal R^{(4)}$ obtained from the analytical solution of the system of Eqs.~(\ref{eq:SystEqsX})-(\ref{eq:SystEqsZ}). Labeled black circles indicate the maximally mixed state (A), the pure product states (B) and the Bell states (C). (b) Zoom into plot (a) in the range $0\leq \mathcal R^{(2)}\leq 1/3^2$ highlighting separable (blue solid lines) and entangled (red dotted lines) Bell diagonal states. Note that for $\mathcal R^{(2)}\leq1/3^3$ all states are separable, and for $1/3^3\leq \mathcal R^{(2)}\leq 1/3^2$ separable and entangled states have a non-zero overlap (striped region). The white dashed curve indicates the entanglement criteria resulting form Eqs.~(\ref{eq:SystEqsX})-(\ref{eq:SystEqsZ}). }
\label{figure_3}
\end{figure}

Direct evaluation of Eq.~(\ref{eq:RandomMoment2}) and (\ref{eq:RandomMoment4}) for Bell diagonal states yields:
\begin{align}
\mathcal R^{(2)}&=\frac{1}{9} (c_x^2 + c_y^2 + c_z^2),\label{eq:Moment2BDS}\\
\mathcal R^{(4)}&=\frac{2}{75} (c_x^4 + c_y^4 + c_z^4)+\frac{27}{25} (\mathcal R^{(2)})^2,\label{eq:Moment4BDS}
\end{align}
which are polynomials of degree smaller or equal than four in the coefficients $c_i$, with $i=x,y,z$. In order to determine the borders of the set of Bell diagonal states in the space spanned by the moments $\mathcal R^{(2)}$ and $\mathcal R^{(4)}$ we have to maximize (minimize), e.g., $\mathcal R^{(4)}$ over the set of states while keeping $\mathcal R^{(2)}$ fixed. The latter optimization has been carried out in Refs.~\cite{MeMoments,MichaelBachelor} and is presented in Fig.~\ref{figure_3}(a). Similarly, we can obtain the borders of the set of separable states by additionally imposing the condition $|c_x|+|c_y|+|c_z|\leq1$. It turns out that solving this optimization problem is equivalent to finding the solution of the system of equations: 
\begin{align}
1&=|c_x|+|c_y|+|c_z|,\label{eq:SystEqsX}\\
\alpha&=c_x^2+c_y^2+c_z^2,\label{eq:SystEqsY}\\
\beta &=c_x^4+c_y^4+c_z^4,\label{eq:SystEqsZ}
\end{align}
where we used the transformations $\alpha:=3^2 \mathcal R^{(2)}$ and $\beta:=15^2 \mathcal R^{(4)}/6-\alpha^2/2$. The relevant solutions of Eqs.~(\ref{eq:SystEqsX})-(\ref{eq:SystEqsZ}) have been derived in Ref.~\cite{MeMoments} and are presented in detail in Fig.~\ref{figure_3}(b). We thus obtain a sufficient entanglement criterion which is a polynomial function of the moments $\mathcal R^{(2)}$ and $\mathcal R^{(4)}$. We note that the latter criterion can be made necessary and sufficient by including the next higher non-vanishing moment $\mathcal R^{(6)}$ \cite{MeMoments,MichaelBachelor}. 

Further on, in the case of general two-qubit states~(\ref{eq:2QuFano}) one can always find corresponding Bell diagonal states $\rho_{\text{BD}}$ having the same moments. On the one hand, this is a consequence of the fact that the moments $\mathcal R^{(t)}$ are LU invariant which allows us to diagonalize the correlation matrix $C=\{c_{i,j}\}_{i,j=x,y,z}$. On the other hand, we can subsequently drop those terms in Eq.~(\ref{eq:2QuFano}) involving only the local Bloch vector components $\boldsymbol a$ and $\boldsymbol b$ by applying a local dephasing operation $\rho\rightarrow \frac{1}{4}(\rho+\sum_{i=x,y,z}\sigma_i\otimes\sigma_i\rho\sigma_i\otimes\sigma_i)$. As the latter transformations are completely positive and, in particular, correspond to the class of LOCC operations, we end up with a resulting Bell diagonal state that has equal moments as the original two-qubit state but whose entanglement cannot have increased. Hence, the derived entanglement criteria for Bell diagonal states yield also sufficient 
entanglement criteria for general two-qubit states. 

Lastly, we note that the results presented in Fig.~\ref{figure_3} can be generalized in order to account for a quantification of entanglement. In Ref.~\cite{MichaelBachelor} it was shown that there exists a one-to-one correspondence between the concurrence of a Bell diagonal state and its corresponding moments $\mathcal R^{(2)}$, $\mathcal R^{(4)}$ and $\mathcal R^{(6)}$. Likewise, for general two-qubit states $\rho$ this one-to-one correspondence turns into a sufficient criterion allowing to lower bound the concurrence of $\rho$.

\subsection{Three qubits} \label{sec:3qubits}

We move on now to the more involved multipartite scenarios and start with the simplest case consisting of three qubits. However, already for three qubits a complete characterization of the set of states in the space spanned by the moments $\mathcal R^{(2)}$ and $\mathcal R^{(4)}$ becomes considerably more difficult due to the increased Hilbert space dimension and the more involved structure of the space of multipartite entangled states. Nevertheless, there have been some formulations of criteria allowing to detect three-qubit entanglement and also genuine three-qubit entanglement based on the second moment only \cite{JulioMarkus,KlocklHuber,NikolaiSectorLengths}. In the following, we will go beyond these results and discuss the characterization of different classes of genuine multipartite entanglement in terms of $\mathcal R^{(2)}$ and $\mathcal R^{(4)}$.

Classes of genuine multipartite entangled states are usually defined through the concept of stochastic local operations and classical communication (SLOCC) \cite{SLOCC3qDuer,SLOCC3qAcin} which form the probabilistic counterpart of  LOCC operations~\cite{NielsenChuang,HorodeckiReview}. In this framework, two pure $N$-qubit states, $\ket\Psi$ and $\ket\Phi$, are equivalent if there exist LOCC operations allowing to transform them into each other with some finite probability. Mathematically, SLOCC equivalence implies the existence of invertible operations $A_i$, with $i=1,\ldots,N$, such that~\cite{SLOCC3qDuer}
\begin{align}
\ket\Psi=A_1\otimes\ldots \otimes A_N \ket\Phi .
\end{align}
The states are then called SLOCC equivalent and the corresponding equivalence classes are referred to as SLOCC classes. For systems consisting of three qubits there exist in total six SLOCC classes: the pure product states $\mathcal S^{(3)}$, three classes of bi-separable states $\mathcal B_{1|23}$, $\mathcal B_{12|3}$ and $\mathcal B_{13|2}$, and two classes of genuinely multipartite entangled states, the $W$- and the GHZ-class \cite{SLOCC3qDuer}, referred to as  $ \mathcal W^{(3)}$ and $ \mathcal{GHZ}^{(3)}$, respectively. As their names suggest, $ \mathcal W^{(3)}$ and $ \mathcal{GHZ}^{(3)}$ consist of those states which are SLOCC equivalent to the states $\ket{W_3}=(\ket{001}+\ket{010}+\ket{100})/\sqrt 3$ and $\ket{\text{GHZ}_3}=(\ket{000}+\ket{111})/\sqrt 2$, respectively.
\begin{figure}[t!]
\begin{center}
\includegraphics[width=0.75\textwidth]{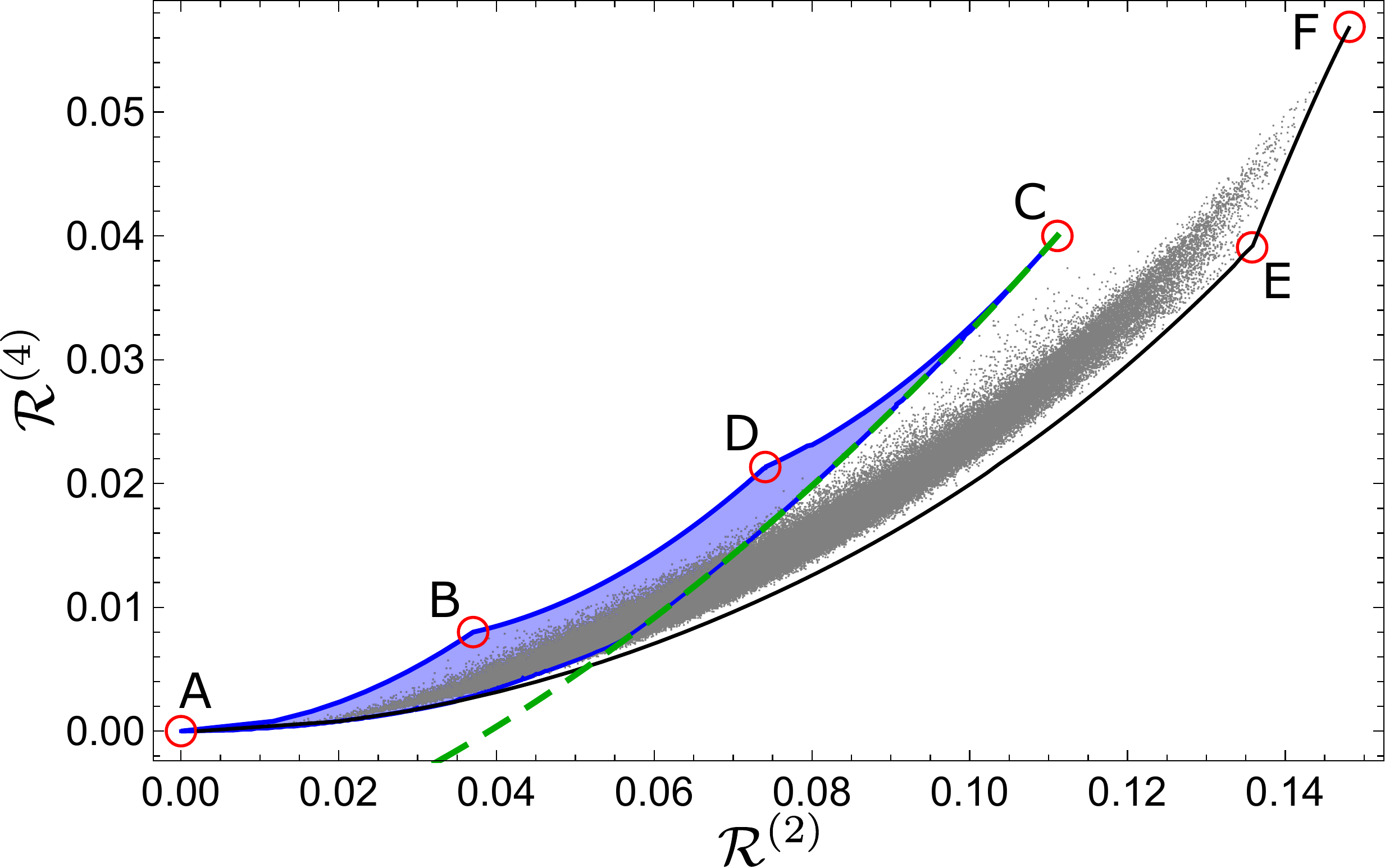}%
\end{center}
\caption{The set of (bi-separable) three-qubit states on the plane spanned by $\mathcal{R}^{(2)}$ and $\mathcal{R}^{(4)}$. The subset of bi-separable states is displayed in light blue with thick blue boundaries. The lower bound of the set of all states is displayed in solid black, and $150,000$ random three-qubit states are displayed in gray. The conjectured criterion in Eq.~(\ref{eq:conj_threequbits}) is displayed as a dashed, green line. The red circles label the maximally mixed state (A), all pure product states (B), bi-separable states of the form $\ket{\phi} \ket{\text{Bell}}$ (C), the uniform mixture  of the bi-separable states $\frac{1}{\sqrt{2}}\ket{0}(\ket{00}+\ket{11})$ and $\frac{1}{\sqrt{2}}\ket{1}(\ket{01}+\ket{10})$ (D), the three-qubit $W$-state (E) and the GHZ-state (F). }
\label{figure_4}
\end{figure}

For mixed states one can define similar classes by taking the convex hull of the corresponding SLOCC classes~\cite{SLOCC3qAcin}. In this way, we obtain the usual set of separable mixed states $\text{Conv}(\mathcal S^{(3)})$, the set of all bi-separable mixed states $\text{Conv}(\mathcal B_{\text{bi-sep}}^{(3)})$, with $\mathcal B_{\text{bi-sep}}^{(3)}:=\mathcal B_{1|23}\cup B_{12|3} \cup \mathcal B_{13|2}$, and consequently $\text{Conv}(\mathcal W^{(3)})$ and $\text{Conv}(\mathcal{GHZ}^{(3)})$, where the convex hull is defined as $\text{Conv}(X):=\{\sum_i p_i x_i| x_i\in X,p_i\geq 0, \sum_i p_i=1\}$. This definition leads to the onion like structure of the set of three-qubit mixed states as presented in Ref.~\cite{SLOCC3qAcin}. We note that in other contexts one might define mixed state analogs of $\mathcal W^{(3)}$ and $ \mathcal{GHZ}^{(3)}$ as $\text{Conv}(\mathcal W^{(3)})\setminus \text{Conv}(\mathcal B_{\text{bi-sep}}^{(3)})$ and $\text{Conv}(\mathcal{GHZ}^{(3)})\setminus \text{Conv}(\mathcal B_{\text{
bi-sep}}^{(3)})$, respectively, in order to ensure that they only contain states which are genuinely multipartite entangled. 
\begin{figure}[t!]
\begin{center}
\includegraphics[width=\textwidth]{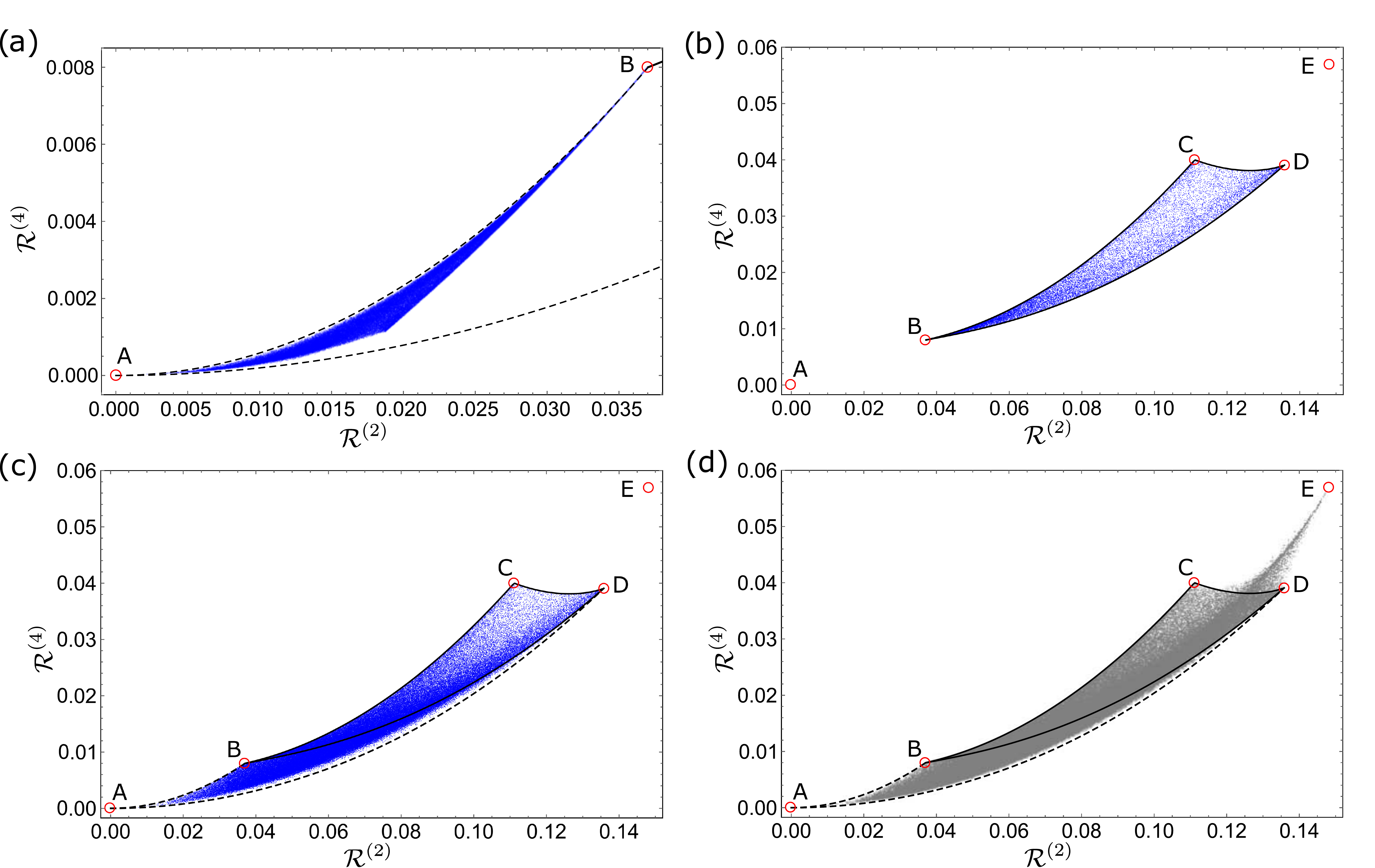}%
\end{center}
\caption{Representations of different three-qubit SLOCC classes in the space spanned by the moments $\mathcal R^{(2)}$ and $\mathcal R^{(4)}$. Plotted points correspond to the moments of randomly generated quantum states from the class of separable states (a),  the pure $W$-class $\mathcal W^{(3)}$ (b), its convex hull $\text{Conv}(\mathcal W^{(3)})$ (c), and  the whole state space of three-qubit density matrices (d). Labeled red circles indicate the maximally mixed state (A), all pure product states (B), bi-separable states of the form $\ket{\phi} \ket{\text{Bell}}$ (C), the three-qubit $W$- (D) and GHZ-state (E) (see App.~\ref{app:RandomStates} for details on the sampling of the random states). While the black solid lines connecting (B), (C) and (D) enclose the set $\mathcal W^{(3)}$, its mixed extension $\text{Conv}(\mathcal W^{(3)})$ is indicated by black dashed lines.}
\label{figure_5}
\end{figure}

As genuine multipartite entanglement constitutes a necessary ingredient for many applications \cite{RaussendorfMBQC, CleveSecret}, we start by distinguishing the set of bi-separable states $\text{Conv}(\mathcal B_{\text{bi-sep}}^{(3)})$ from the set of genuinely multipartite entangled states in $\text{Conv}(\mathcal{GHZ}^{(3)})\setminus \text{Conv}(\mathcal B_{\text{bi-sep}}^{(3)})$ and $\text{Conv}(\mathcal W^{(3)})\setminus \text{Conv}(\mathcal B_{\text{bi-sep}}^{(3)})$ in order to find new entanglement criteria based on higher moments. To that end, we numerically optimize the moments $\mathcal R^{(2)}$ and $\mathcal R^{(4)}$ for mixed bi-separable states and compare the result to an optimization over the set of all states. The results are depicted in Figure~\ref{figure_4} and show a clear difference between the two sets, allowing for improved entanglement detection compared to any criterion based on  $\mathcal R^{(2)}$ alone. The boundary between the two sets can be approximated by a quadratic function, 
which is displayed in the figure as well. This yields the following conjectured entanglement criterion for three-qubit states:
\begin{equation}\label{eq:conj_threequbits}
    \mathcal{R}^{(4)}_{\rho_\text{bisep}} \geq \frac{1}{425}[972(\mathcal{R}^{(2)}_{\rho_\text{bisep}})^2 + 90 \mathcal{R}^{(2)} - 5].
\end{equation}
We note that it is possible to prove a similar bound on the set of pure bi-separable states. To do so, one simply assumes an arbitrary product of single- and two-qubit states $\ket{\text{Bisep}}=\ket{\phi}\ket{\Psi}$ for which the respective fourth moment factorizes to $\mathcal{R}^{(4)}_{\ket{\text{Bisep}}}=\frac{1}{5}\mathcal{R}^{(4)}_{\ket\Psi}$. Subsequently, one can apply the results of Sec.~\ref{sec:2QubitStuff} in order to obtain the criterion $\mathcal{R}^{(4)}_{\ket{\text{Bisep}}}\geq 5^4 \mathcal R^{(2)}_{\ket{\text{Bisep}}}/3^5$. The latter leads to a slightly tighter curve than the one corresponding to Eq.~(\ref{eq:conj_threequbits}), shown in Fig.~\ref{figure_4}, thus confirming the conjecture for pure separable states.

Next, we focus on the remaining SLOCC classes of fully separable states and the two genuinely multipartite entangled ones.
Figure~\ref{figure_5} contains the results of a numerical analysis aiming at identifying these classes through a mixture of numerical optimizations and the generation of random states. For instance, Fig.~\ref{figure_5}(b) shows a numerical estimation of the borders of the pure $W$-class $\mathcal W^{(3)}$, which is confirmed by more than $10^5$ randomly generated three-qubit states. Both, the numerical estimation of the borders of the pure $W$-class and the generation of random states is based on the three-qubit standard form presented in Ref.~\cite{StForm3qAcin}. The latter allows us to numerically determine the border of  $\mathcal W^{(3)}$ by numerically optimizing the moments with respect to the standard form parameters. Furthermore, by drawing the same parameters randomly we were able to sample states from $\mathcal W^{(3)}$ (see App.~\ref{app:RandomStates} for details). As the moments are invariant under LU transformations, this suffices to sample exhaustively points in the $(\mathcal R^{(2)},\mathcal R^{(4)})$-plane corresponding to states in $\mathcal W^{(3)}$.

Further on, in Fig.~\ref{figure_5}(a), (c) and (d) we focus on the mixed SLOCC classes $\text{Conv}(\mathcal S^{(3)})$, $\text{Conv}(\mathcal W^{(3)})$ and $\text{Conv}(\mathcal{GHZ}^{(3)})$, respectively. However, due to the structure of the convex hull these classes are more difficult to characterize. We thus only estimated their boundaries roughly by minimizing the moments over a subset of permutational invariant mixed states: $\text{Conv}(\{\ketbra{000},\ketbra{W_3},\mathbbm 1_8/2^3\})\subset \text{Conv}(\mathcal W^{(3)})$, the results of which are indicated in Fig.~\ref{figure_5} by black dashed lines. Moreover, by sampling states from $\mathcal S^{(3)}$, $\mathcal W^{(3)}$ and $\mathcal{GHZ}^{(3)}$, applying random LU transformations and subsequently mixing them with randomly drawn mixing parameters, we are able to sample states from the corresponding mixed classes. The latter seem to cover quiet well the $(\mathcal R^{(2)},\mathcal R^{(4)})$-plane as presented in Fig.~\ref{figure_5}. However, we stress that some areas of the $(\mathcal R^{(2)},\mathcal R^{(4)})$-plane are underrepresented by this procedure because of the low sampling probability of the corresponding states. For instance, one can find sequences of states in $\text{Conv}(\mathcal W^{(3)})$ which converge towards the uniform mixture of the bi-separable states $\frac{1}{\sqrt{2}}\ket{0}(\ket{00}+\ket{11})$ and $\frac{1}{\sqrt{2}}\ket{1}(\ket{01}+\ket{10})$ (see also Fig.~\ref{figure_4}). Hence, it is expected that the upper boundary of the set $\text{Conv}(\mathcal W^{(3)})$ shows a similar structure as that of the mixed bi-separable sates presented in Fig.~\ref{figure_4}.

In conclusion, the above analysis shows that a discrimination of the mixed SLOCC classes $\text{Conv}(\mathcal S^{(3)})$, $\text{Conv}(\mathcal W^{(3)})$ and $\text{Conv}(\mathcal{GHZ}^{(3)})$ in terms of $\mathcal R^{(2)}$ and $\mathcal R^{(4)}$ is possible.


\subsection{Four qubits} \label{sec:4qubits}
In this section we carry out a similar analysis in the slightly more complicate case consisting of four qubits. The main aim will be to give the reader an impression of the growing complexity of the problem of characterizing SLOCC classes in the space spanned by the moments $\mathcal R^{(2)}$ and $\mathcal R^{(4)}$  by discussing important differences to the results presented in Sec.~\ref{sec:3qubits}. 
\begin{figure}[t!]
\begin{center}
\includegraphics[width=\textwidth]{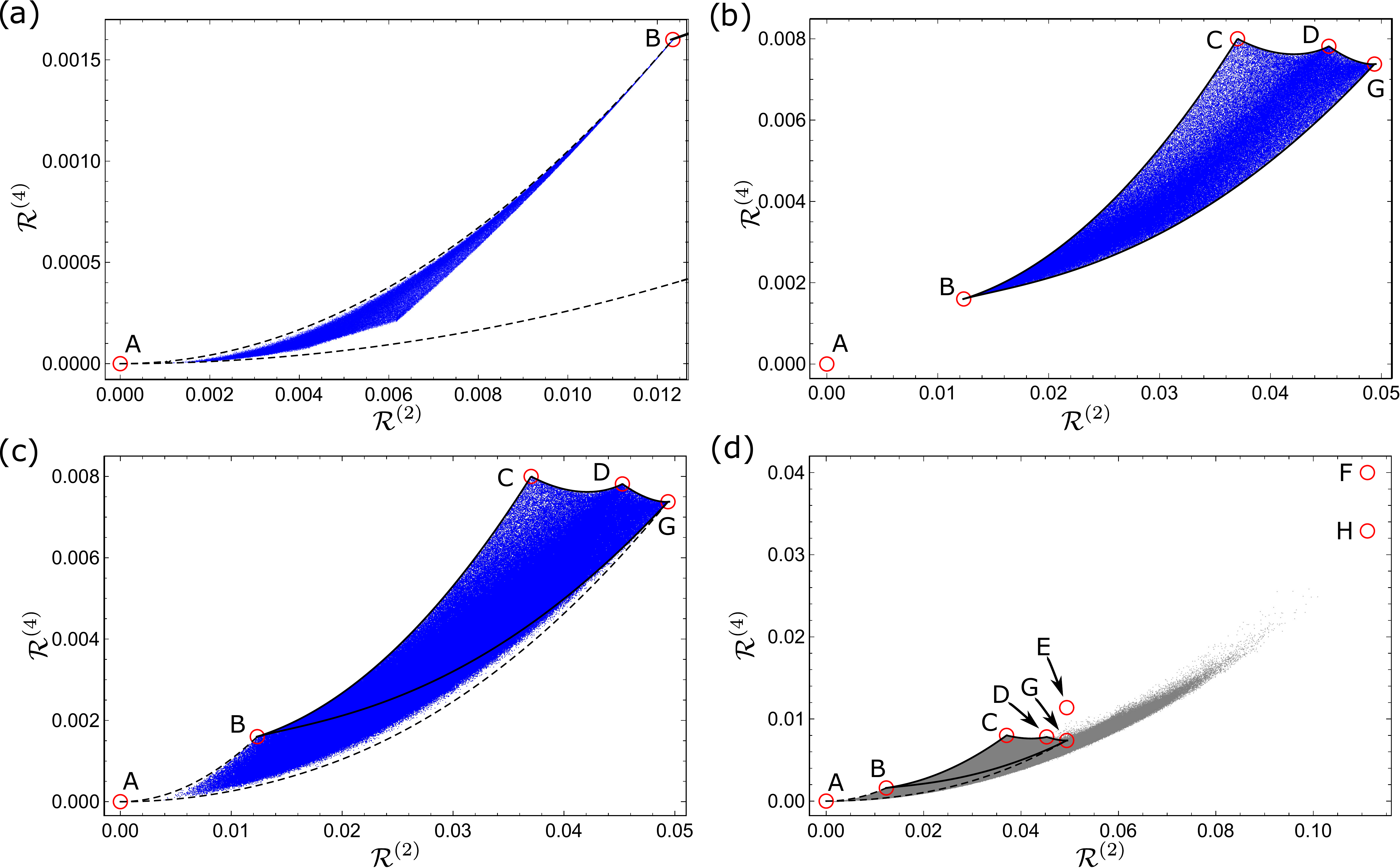}%
\end{center}
\caption{Representations of different four-qubit SLOCC classes in the space spanned by the moments $\mathcal R^{(2)}$ and $\mathcal R^{(4)}$. Plotted points correspond to the moments of randomly generated quantum states from the class of separable states (a), the pure $W$-class $\mathcal W^{(4)}$ (b), its convex hull $\text{Conv}(\mathcal W^{(4)})$ (c), and the whole state space of four-qubit density matrices (d). Labeled red circles indicate the maximally mixed state (A), all pure product states (B), tri-separable states of the form $\ket{\phi}\ket{\phi'} \ket{\text{Bell}}$ (C), bi-separable states of the form $\ket{\phi}\ket{W}$ (D), $\ket{\phi}\ket{\text{GHZ}_3}$ (E) and $\ket{\text{Bell}}\ket{\text{Bell}}$ (F), and the four-qubit $W$- (G) and GHZ-state (H) (see App.~\ref{app:RandomStates} for details on the sampling of the random states). Again, the black solid lines enclose the set $\mathcal W^{(4)}$ and its mixed extension $\text{Conv}(\mathcal W^{(4)})$ is indicated by black dashed lines.}
\label{figure_6}
\end{figure}

Figure~\ref{figure_6} shows an overview of the characterization of four-qubit SLOCC classes. From this we first note the remarkable resemblance between the set of separable three- and four-qubit states (see Fig. \ref{figure_5}(a) and \ref{figure_6}(a)). In contrast, the respective four-qubit $W$-classes, $\mathcal W^{(4)}$ and $\text{Conv}(\mathcal W^{(4)})$,  show some more detailed structures which correspond to different types of bi- and tri-separable states (see Fig.~\ref{figure_6}(b) and (c)). Furthermore, one recognizes that in the four-qubit case the relative size of the pure, as well as the mixed $W$-classes is smaller in comparison to the set of all states (see Fig.~\ref{figure_6}(d)). Lastly, we see that $\mathcal R^{(2)}$ is maximized simultaneously by the four qubit GHZ state $\ket{\text{GHZ}_4}$ and the bi-separable state 
$\ket{\text{Bell}}\ket{\text{Bell}}$ consisting of a product of two Bell states, the latter of which even maximizes $\mathcal R^{(4)}$. This last point shows that it is no longer possible to detect genuine multipartite entanglement based solely on $\mathcal R^{(2)}$, as its maximum within the set $\text{Conv}(\mathcal B^{(4)}_\text{bi-sep})$ coincides with its maximum over all four-qubit states. A similar conclusion can be reached for $\mathcal R^{(4)}$, as its value for the state $\ket{\text{Bell}}\ket{\text{Bell}}$ is larger than that for $\ket{\text{GHZ}_4}$. If the detection of genuine multipartite entanglement is possible for more than three qubits by suitably combining moments of different orders will be subject of future investigations.

In order to understand some of the above observations, it is important to note that the structure of multipartite entanglement classes of four qubits is already considerably more complicated than for three qubits. In fact, in the case of four qubits we are dealing with infinitely many SLOCC  equivalence classes~\cite{Conny4qu}. Moreover, the onion like structure of the mixed SLOCC classes given in the case of three qubits is no longer present~\cite{StForm3qAcin,SLOCC3qAcin}. Hence, it is generally possible to find states that are not contained in the $W$-class and at the same time are not genuinely multipartite entangled. 

In the next section we will push further in this direction and investigate the discrimination of $W$-class states for an arbitrary number of qubits.

\section{\label{eq:F}Discrimination of $W$-class states}\label{sec:DiscWclass}

In this Section we derive a criterion that allows for a discrimination of mixed $W$-class states for an arbitrary number $N$ of qubits  based on the second moment $\mathcal R^{(2)}$. In particular, we find that
\begin{align}
\mathcal R^{(2)}_{\rho}\leq \frac{5-\frac{4}{N}}{3^N}=:\chi^{(N)}, 
\label{eq:WclassBound}
\end{align}
for all $\rho\in \text{Conv}(\mathcal W^{(N)})$, with equality for the pure $W$-state
\begin{align}
\ket{W_N}=\frac{1}{\sqrt N}\left(\ket{10\ldots0}+\ket{010\ldots0}+\ldots + \ket{0\ldots01}\right).
\label{eq:NquWstate}
\end{align}
Hence, any multi-qubit state whose second moment violates the criterion~(\ref{eq:WclassBound}) cannot belong to the mixed $W$-class. While we have addressed such an indirect characterization of the class $ \text{Conv}(\mathcal W^{(N)})$ already by numerical means for up to ten qubits in Ref.~\cite{MeMoments}, we provide here an analytical proof of the criterion~(\ref{eq:WclassBound}) for arbitrary $N$. The basic proof idea is to maximize $\mathcal R^{(2)}$ first over all pure states contained in $\mathcal W^{(N)}$ and then conclude by convexity that the same value also yields the maximum over the class $ \text{Conv}(\mathcal W^{(N)})$. Furthermore, the maximum of $\mathcal R_\rho^{(2)}$, with $\rho \in \mathcal W^{(N)}$, can be determined via a rather simple geometric argument, as the value of $\mathcal R_\rho^{(2)}$ can be associated with a section of the surface area of a square with side length $1$ (see Fig.~\ref{app:GeomProof}). For further details on the proof of criterion~(\ref{eq:WclassBound}) see App.
~\ref{app:ProofBound}.
\begin{figure}[t!]
\begin{center}
\includegraphics[width=0.7\textwidth]{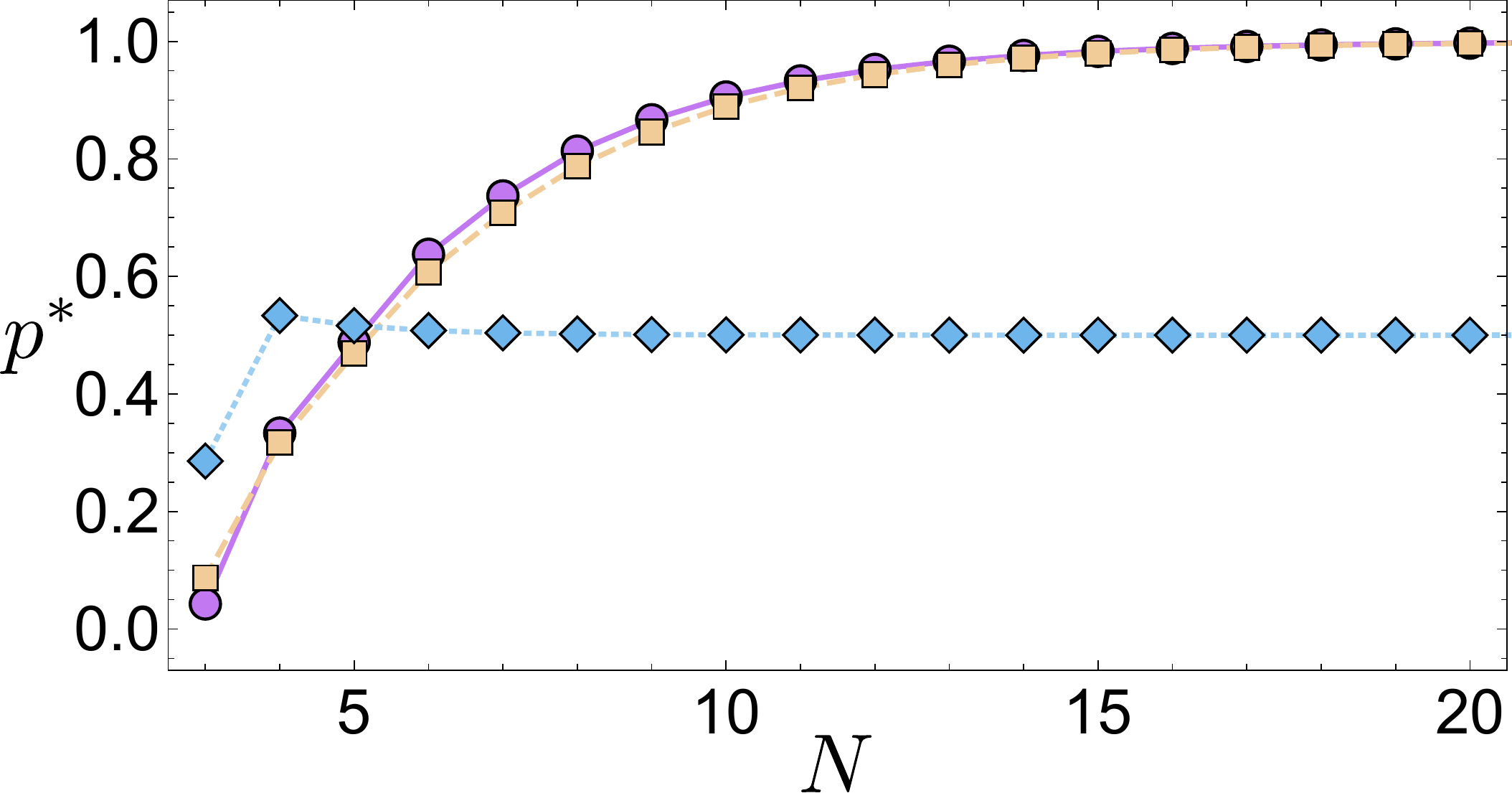}%
\end{center}
\caption{Plot of the threshold value $p^*$ for the detection of the noisy GHZ state $\rho_\text{GHZ}(p,N)$  in terms of the criteria $\mathcal R^{(2)}\leq \chi^{(N)}$ (purple circles), $\text{Lin}(\mathcal R^{(2)},\mathcal R^{(4)})\leq 0$ (yellow squares) and $\expec{W^{(N)}_{\text{GHZ}}}\geq 0$ (cyan diamonds), respectively, as a function of the number $N$ of qubits.  } 
\label{figure_7}
\end{figure}

In order to investigate the performance of criterion~(\ref{eq:WclassBound}) we apply it to a $N$-qubit GHZ state mixed with white noise, i.e., $\rho_\text{GHZ}(p,N):=p \mathbbm 1/2^N+(1-p) \ketbra{\text{GHZ}_N}$, where $\ket{\text{GHZ}_N}=\big(\ket{0}^{\otimes N}+\ket{1}^{\otimes N}\big)/\sqrt 2$. As $\mathcal R^{(2)}_{\mathbbm 1/2^N}=0$, for all $N$, it is easy to see that the second moment of the noisy GHZ-state $\rho_\text{GHZ}(p,N)$ is given by
\begin{align}
\mathcal R^{(2)}= \mathcal R^{(2)}_{\ket{\text{GHZ}_N}} (1-p)^2,
\label{eq:R2NoisyGHZ}
\end{align}
where 
\begin{align}
 \mathcal R^{(2)}_{\ket{\text{GHZ}_N}}= \begin{cases}
2^{N-1}/3^N, &N\ \mathrm{odd}, \\
(2^{N-1}+1)/3^N,  &N\ \mathrm{even}.
\end{cases}
\label{eq:R2GHZ}
\end{align}
In particular, it has been proven recently in Refs.~\cite{tran2,JensR2max,NikolaiSectorLengths} that the GHZ-value~(\ref{eq:R2GHZ}) also yields the maximum of the second moment $\mathcal R^{(2)}$. Hence, the noisy GHZ state $\rho_\text{GHZ}(p,N)$ approaches the minimum and maximum of $\mathcal R^{(t)}$ for the maximally mixed state and the GHZ state, respectively. With Eq.~(\ref{eq:R2NoisyGHZ}) in hand we can easily determine the noise threshold $p^*$ up to which the GHZ state violates the criterion~(\ref{eq:WclassBound}): 
\begin{align}
p^*= \begin{cases}
1-2^{\frac{1}{2}-\frac{N}{2}} \sqrt{5-\frac{4}{N}}, &N\ \mathrm{odd}, \\
1-\frac{\sqrt{10 N-8}}{\sqrt{\left(2^{N}+2\right) N}},  &N\ \mathrm{even}.
\end{cases}
\label{eq:Wpstar}
\end{align}
As comparison, we also calculate the threshold value $\tilde p^*$ obtained from the SLOCC witness found in Refs.~\cite{SLOCC3qAcin,SLOCCwitSolano,SLOCCwitConny}. The latter reads $W^{(N)}_{\text{GHZ}}:=\lambda \mathbbm 1_{2^N}-\ketbra{\text{GHZ}_N}$, with $\lambda=3/4$ for $N=3$ and $\lambda=1/2$ for $N\geq 4$, and leads to
\begin{align}
\tilde p^*= \begin{cases}
\frac{2}{7}, &N=3, \\
\frac{2^{N-1}}{2^N-1},  &N\geq 4.
\end{cases}
\label{eq:ConnyWitp}
\end{align}
Equations~(\ref{eq:Wpstar}) and (\ref{eq:ConnyWitp}) are plotted in Fig.~\ref{figure_7}. While the SLOCC witness is more noise robust for low number of qubits, our criterion~(\ref{eq:WclassBound}) performs better for $N> 5$. In this respect, one has to keep in mind that the SLOCC witness $W^{(N)}_{\text{GHZ}}$ also detects the genuine multipartite entanglement of the states which partly explains the discrepancies between the performances given in Eqs.~(\ref{eq:Wpstar}) and (\ref{eq:ConnyWitp})~\cite{OtfriedReview}.  Asymptotically, for $N\rightarrow \infty$, we find that the threshold value $p^*$ approaches $1$ which is a direct consequence of the fact that the bound~(\ref{eq:WclassBound}) approaches zero in the same limit. Equation~(\ref{eq:Wpstar}) thus demonstrates the decreasing relevance of the $W$-class in large multipartite systems.

Hence, if one is interested in the $W$-class alone it is already quiet exhaustive to consider only the second moment $\mathcal R^{(2)}$. We further investigate this issue by considering a simple generalizaiton of the criterion (\ref{eq:WclassBound}) which includes also the fourth moment $\mathcal R^{(4)}$. The idea is to define a linear criterion in the $(\mathcal R^{(2)},\mathcal R^{(4)})$-plane that passes through the $W$-state $\ket{W_N}$ and the bi-seprable state $\ket\phi\ket{W_{N-1}}$. In App.~\ref{app:NikWit} we prove that this line indeed forms another possible criterion, denoted as $\text{Lin}(\mathcal R^{(2)},\mathcal R^{(4)})\leq 0$, allowing to discriminate states of the $W$-class $\text{Conv}(\mathcal W^{(N)})$. As for Eq.~(\ref{eq:WclassBound}) we further study the performance of this criterion by applying it to the noisy GHZ-state $\rho_\text{GHZ}(p,N)$ and include the results into Fig.~\ref{figure_7}. We find that in this particular case the generalized criterion $\text{Lin}(\mathcal R^{(2)},\mathcal R^{(4)})\leq 0$ does  not provide an advantage over that of Eq.~(\ref{eq:WclassBound}). Nevertheless, it allows one to detect other classes of states which are not in reach of criterion~(\ref{eq:WclassBound}). For instance, the states $\ket 0^{\otimes (N-3)}\ket{\text{GHZ}_3}$ yield the second moment $\mathcal R^{(2)}=4/3^N$  which is smaller or equal to the bound $\chi^{(N)}$, for all $N\geq 4$ (see Fig.~\ref{figure_6}(d) for the case $N=4$). However, it violates the generalized criterion $\text{Lin}(\mathcal R^{(2)},\mathcal R^{(4)})\leq 0$, for all $N\geq 3$. Hence, depending on the class of states under consideration, including the fourth moment can provide an advantage in detecting states outside of the W-class $\text{Conv}(\mathcal W^{(N)})$.

\section{Conclusion}\label{sec:Conclusion}


In conclusion, we showed how to characterize multipartite entangled states in a reference frame independent manner through moments of correlation functions obtained from locally randomized measurements. As the latter correspond mathematically to uniform averages over polynomials of  finite degree, we were able to exploit the concept of pseudo-random processes in order to  evaluate them exactly for variable system sizes. In particular, we showed, in the case of qubit systems, how to express the moments in terms of specific examples of spherical designs and discussed the dependency of the involved number of measurements on the respective order of the moments. As application of the  introduced framework, we presented an analysis of the structure of different classes of multipartite  entangled states of few qubit systems using the first two  non-vanishing moments. Lastly, we introduced two novel criteria that allow for a discrimination of $W$-type entangled states in terms of the second and fourth moment for an 
arbitrary number of qubits. 

The presented framework for the characterization of multipartite entanglement based on randomized measurements promises advantages, in particular, in the limit of large system sizes where conventional methods, such as state tomography, become impractical. In this limit one can attempt to evaluate the moments~(\ref{eq:RandomMomentsQudits}) approximately using a finite number of randomly measured correlation functions. As such an approach will consequently lead to additional statistical errors, it will be subject of future investigations to develop analytical methods allowing to predict such errors in advance.


\section*{Acknowledgements}
We thank Cornelia Spee and Jasmin Meinecke for fruitful discussions. This work was supported by the ERC (Consolidator Grant 683107/TempoQ), and the DFG. NW acknowledges support by the QuantERA grant QuICHE and the BMBF. AK acknowledges support by the Georg H. Endress foundation. The article processing charge was funded by the Baden-Wuerttemberg Ministry of Science, Research and Art and the University of Freiburg in the funding programme Open Access Publishing.
\begin{appendix}

\section{Qubit unitary $5$-design $SL(2,\mathbb F_5)$}\label{app:Unitary5Design}

First, we extract a set of generators of $SL(2,\mathbb F_5)$ from the GAP character library using the package REPSN, as outlined in \cite{Gross5Design}, leading to:
\begin{align}
&\begin{pmatrix}
-1 & 0 \\
0 & -1 
\end{pmatrix},
\begin{pmatrix}
-\omega^{11} - \omega^{14} & 
  \omega^6 + \omega^9 \\
  -\omega - \omega^2 - \omega^4 - 
   \omega^7 - \omega^8 - \omega^{13} & 
  \omega^{11} + \omega^{14}
\end{pmatrix},\nonumber \\
&\begin{pmatrix}
\omega^{10} &\omega^{11} + \omega^{14}\\
-\omega^2 - 
    \omega^8 & -\omega^{10} 
\end{pmatrix},
\begin{pmatrix}
0& \omega^5\\
-\omega^{10} & -\omega^3 - \omega^{17} 
\end{pmatrix},
\label{eq:Generators5Design}
\end{align}
with $\omega=e^{i 2\pi/15}$. 
Next, with the matrices (\ref{eq:Generators5Design}) we generate the $120$ group elements of $SL(2,\mathbb F_5)=\{S_k|k\in\{1,\ldots,120\}\}$ from which we can then extract an appropriate unitary representation using the transformation:
\begin{align}
U_k=\sqrt{P} S_k \sqrt{P}^{-1},\ \ \text{with}\ \ P:=\sum_{k=1}^{120} S_k^\dagger S_k>0.
\end{align}
Further on, after eliminating those matrices that are equal up to a global phase $e^{i\phi}$, we end up with a set of $60$ unitary matrices representing the corresponding unitary $5$-design.

\section{Generation of random states}\label{app:RandomStates}

In Sec.~\ref{sec:3qubits} and \ref{sec:4qubits} we used standard forms of multi-qubit states in order to sample random states from different SLOCC classes. Standard forms are multipartite generalizations of the Schmidt decomposition, e.g., a three qubit standard form was introduced in \cite{StForm3qAcin}:
\begin{align}
\ket{\Psi_{\text{s.f.}}}=\lambda_0 \ket{000}+\lambda_1e^{i\phi}\ket{100}+\lambda_2 \ket{101}+\lambda_3\ket{110}+\lambda_4\ket{111},
\label{eq:3qubitStForm}
\end{align}
where $\lambda_i\geq0$, with $i=0,\ldots,4$, and $0\leq \phi\leq \pi$. Any pure three-qubit state can be represented by Eq.~(\ref{eq:3qubitStForm}) plus an appropriate LU transformation. For a generalization of Eq.~(\ref{eq:3qubitStForm}) to $N$-partite systems of local dimensions $d$ we refer the reader to Ref.~\cite{StFormNq}. More generally, standard forms allow for a characterization of different SLOCC classes. For instance, with $\lambda_4=\phi=0$ and $\lambda_i>0$, for $i=0,1,2,3$, Eq.~(\ref{eq:3qubitStForm}) becomes a standard form describing the three-qubit $W$-class $\mathcal W^{(3)}$. A similar standard form for the $N$-qubit $W$-class $\mathcal W^{(N)}$ has been presented in Ref.~\cite{StFormWsts}:
\begin{align}
\ket{W_N(\vec x)}=\sqrt{x_0} \ket{0\ldots0}+\sqrt{x_1}\ket{10\ldots0}+\ldots + \sqrt{x_{N-1}}\ket{0\ldots010}+\sqrt{x_{N}}\ket{0\ldots01}.
\label{app:Wstandardform}
\end{align}
where $\vec x=(x_0,\ldots,x_{N})^\text{T}$, and $|\vec x|_1={\sum_{i=0}^{N} x_i}=1$. 
Hence, we can sample random pure states from each of the mentioned classes by drawing randomly the respective real parameters and apply additional random LU transformations. The latter step can be skipped in our case as we are interested in the distribution respective moments $\mathcal R^{(t)}$ of the states which are LU invariant. In Fig.~\ref{figure_5}(b) and \ref{figure_6}(b) of the main text we present randomly sampled three- and four-qubit $W$-states in the $(\mathcal R^{(2)},\mathcal R^{(4)})$-plane. 

In order to sample mixed quantum states different strategies have been studied in the literature. One way is to draw a random pure state from a higher dimensional Hilbert space and subsequently trace over the extended dimensions~\cite{SamplingSurvey}. This approach leads to random states which are not homogeneously enough distributed in the space spanned by the moments $\mathcal R^{(2)}$ and $\mathcal R^{(4)}$. Alternatively, we can use the above methods for sampling random pure states, apply additional random LU transformations and subsequently mix them with randomly drawn mixing parameters, e.g., for $\text{Conv}(\mathcal W^{(N)})$ we sample $\rho_W=\sum_{\alpha=1}^{2^N} p_\alpha \left(\bigotimes_{i=1}^N U_i^{(\alpha)}\right)\ketbra{W_N(\vec x_\alpha)}\left(\bigotimes_{i=1}^N {U_i^{(\alpha)}}\right)^\dagger$, with randomly drawn $x_i^{(\alpha)}$'s and $p_\alpha$'s with $\sum_{\alpha=1}^{2^N} p_\alpha=1$. Here, in contrast to the case of pure states, we cannot simply skip the step of applying additional random LU transformations as the process of mixing the individual pure states can in general lead to a larger class of mixed states. However, we 
found numerically that mixing of pure standard form states already exhausts quite well the numerically estimated lower boundaries shown in Fig.~\ref{figure_5} and \ref{figure_6}. Nevertheless, the outlined methods to sample mixed states have of course their limits, as explained in the main text.

\section{Proof of the criteria presented in Sec.~\ref{sec:DiscWclass}}\label{app:ProofBound}
\subsection{Criterion based on $\mathcal R^{(2)}$}
In the following, we prove the statement:
\begin{align}
\mathcal R^{(2)}_{\rho}\leq \frac{5-\frac{4}{N}}{3^N}:=\chi^{(N)}, 
\label{app:WclassBound}
\end{align}
for all $\rho\in \text{Conv}(\mathcal W^{(N)})$, with equality for the pure $W$-state~(\ref{eq:NquWstate}). 

\begin{proof}
In order to prove the bound presented in Eq.~(\ref{eq:WclassBound}) we first have to determine the maximum value of $\mathcal R^{(2)}$ for all pure $W$-states, i.e., $\chi^{(N)}:=\max_{\ket{\Psi} \in \mathcal W^{(N)}}  {\mathcal R^{(2)}_{\ket{\Psi}}}$. Given $\chi^{(N)}$, we can subsequently conclude by convexity of $\mathcal R^{(2)}$ that for all $\rho_{\mathcal W}\in \text{Conv}(\mathcal W^{(N)})$ the following inequality holds:
\begin{align}
\mathcal R^{(2)}_{\rho_{W}}\leq \sum_\alpha p_\alpha \mathcal R^{(2)}_{\ket{W_N}} \leq  \max\limits_{\rho'\in \mathcal W^{(N)}} \mathcal R^{(2)}_{\ket{W_N}}:=\chi^{(N)}.
\label{eq:GHZcritR2}
\end{align}
\begin{align}
\mathcal R^{(2)}_{\ket{W_N}}=\frac{1}{3^N}\left[1+8\sum_{i<j; i,j=1}^{N}1/N^2\right]=\frac{1}{3^N}\left(5-\frac{4}{N}\right).
\label{app:R2WstMax}
\end{align}
\begin{figure}[t!]
\begin{center}
\includegraphics[width=\textwidth]{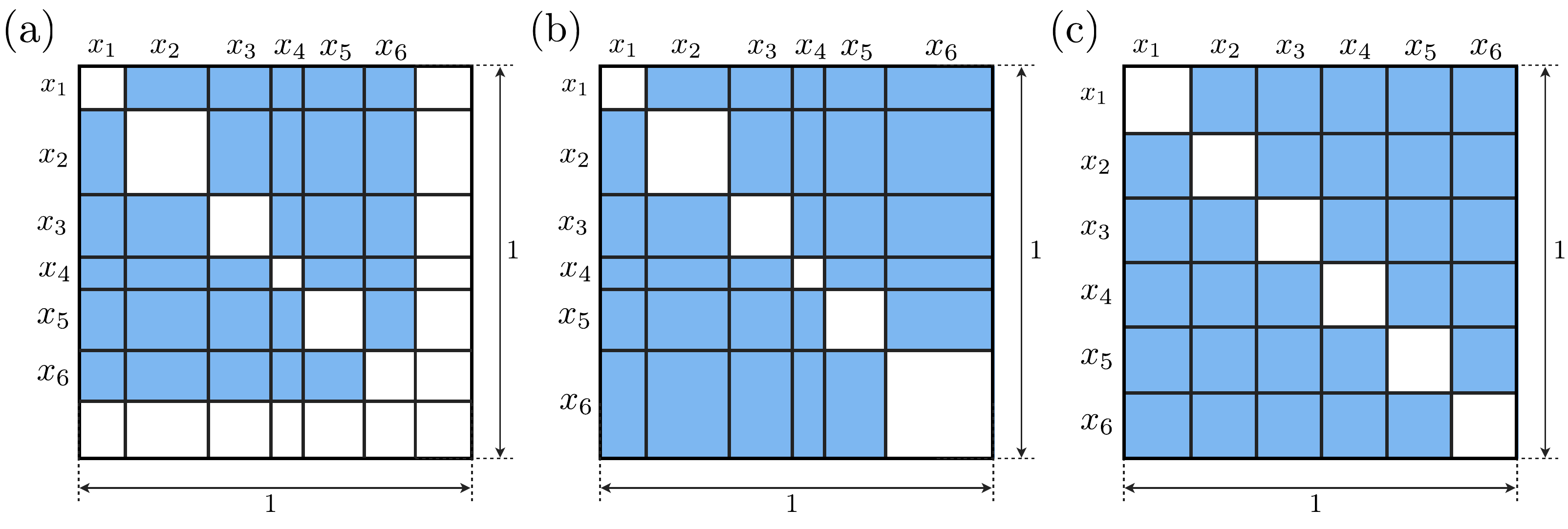}%
\end{center}
\caption{Geometric representation of the value of $\mathcal R^{(2)}_{\ket{W_N(\underline{\vec x})}}$, for $N=6$, as the blue colored area within a square of side length $1$ for three different cases: with $\sum_{i=1}^{N}x_i<1$ (a), with $\sum_{i=1}^{N}x_i=1$ (b), and with maximal $\mathcal R^{(2)}_{\ket{W_N(\underline{\vec x})}}$ (c).   } 
\label{app:GeomProof}
\end{figure}

In order to calculate the maximum $\chi^{(N)}$, we first evaluate $\mathcal R^{(2)}$ with respect to the standard form~(\ref{app:Wstandardform}) and subsequently optimize it with respect to the involved parameters $(x_0,\ldots,x_N)$. 
This is sufficient because, on the one hand, for every state $\ket\Psi \in \mathcal W^{(N)}$ one can find an appropriate vector $\vec x'$ and a LU transformation $U_1'\otimes \ldots \otimes U_N'$ such that $\ket\Psi=(U_1'\otimes \ldots \otimes U_N')\ket{W_N(\vec x^*)}$ and, on the other hand, the moments $\mathcal R^{(t)}$ are invariant under LU transformations. Evaluating $\mathcal R^{(t)}$ in terms of Eq.~(\ref{app:Wstandardform}) thus yields
\begin{align}
\mathcal R^{(2)}_{\ket{W_N(\vec x)}}=\frac{1}{3^N}\left[|\vec x|_2^2+2x_0\sum_{i=1}^{N} x_i+10\sum_{i<j; i,j=1}^{N}x_ix_j\right],
\label{app:R2Wstandardform1}
\end{align}
where $\vec x=(x_0,\ldots,x_{N})^\text{T}$, and $|\vec x|_2=\sqrt{\sum_{i=0}^{N} x_i^2}$. Next, we substitute the constraint $x_0=1-\sum_{i=1}^{N}x_i$, yielding
\begin{align}
\mathcal R^{(2)}_{\ket{W_N(\underline{\vec x})}}=\frac{1}{3^N}\left[1+8\sum_{i<j; i,j=1}^{N}x_ix_j\right],
\label{app:R2Wstandardform2}
\end{align}
where $\underline{\vec x}=(x_1,\ldots,x_{N})^\text{T}$. From here it remains to be checked if the maximum of Eq.~(\ref{app:R2Wstandardform2}) is contained in the interior of the parameter space or lies on its boundaries determined by the constraints $1\geq \sum_{i=1}^{N}x_i$, and $x_i\geq 0$, for all $i=1,\ldots,N$. This question can already be answered in terms of a simple geometric picture. To see this, we rewrite Eq.~(\ref{app:R2Wstandardform2}) as follows:
\begin{align}
\mathcal R^{(2)}_{\ket{W_N(\underline{\vec x})}}=\frac{1}{3^N}\left\{1+4\left[\left(\sum_{i=1}^{N}x_i\right)^2-\sum_{i=1}^N x_i^2\right]\right\},
\label{app:R2Wstandardform3}
\end{align}
and note that the $x_i$-dependent part in the rectangular brackets determines the size of the blue area of the square with length $1$ presented in Fig.~\ref{app:GeomProof}. From this geometric picture one recognizes that in order to reach the maximum value of $\mathcal R^{(2)}_{\ket{W_N(\underline{\vec x})}}$, the vector $\underline{\vec x}$ has be on the boundary determined by $\sum_{i=1}^{N}x_i=1$. This  is clear because if one takes an exemplary value of $\underline{\vec x}$ with $\sum_{i=1}^{N}x_i<1$, as shown in Fig.~\ref{app:GeomProof}(a), one can simply increase the value of $\mathcal R^{(2)}_{\ket{W_N(\underline{\vec x})}}$ by increasing one of the components $x_i$. Hence, if all the components $x_i$ are non-zero and the maximum is on the boundary $\sum_{i=1}^{N}x_i=1$, we are left with 
\begin{align}
\mathcal R^{(2)}_{\ket{W_N(\underline{\vec x})}}=\frac{1}{3^N}\left\{1+4\left[1-\sum_{i=1}^N x_i^2\right]\right\},
\label{app:R2Wstandardform4}
\end{align}
which is maximal whenever $\sum_{i=1}^N x_i^2$ is minimal, i.e., if $x_i=1/N$ for all $i=1,\ldots,N$ (see Fig.~\ref{app:GeomProof}(c)). This solution corresponds exactly to the $N$-qubit $W$-state $\ket{W_N}$ and leads to the following value of the second moment

In order to prove that Eq.~(\ref{app:R2WstMax}) is the true maximum, it remains to be shown that $\mathcal R^{(2)}$ takes no larger value on the other boundaries. The remaining boundaries are those where $1= \sum_{i=1}^{N}\underline x_i$ and simultaneously $k$ of the $N$ remaining $\underline x_i$, with $i=1,\ldots,k$, and all permutations thereof, are equal to zero. In all those cases we can use a similar geometric argument to find that the maxima are reached whenever the remaining $N-k$ components read $x_i=1/(N-k)$, leading to 
\begin{align}
\mathcal R^{(2)}_{\ket{W_N({\underline{\vec x}(k)})}}=\frac{1}{3^N}\left[1+8\sum_{i<j; i,j=1+k}^{N}\frac{1}{(N-k)^2}\right]=\frac{1}{3^N}\left(5-\frac{4}{N-k}\right).
\label{app:R2BisepWstMax}
\end{align}
where $\underline{\vec x}(k)$ denote all vectors with $k$ vanishing components and $N-k$ components equal to $1/(N-k)$. Hence, we have $\mathcal R^{(2)}_{\ket{W_N({\underline{\vec x}(k)})}}<\mathcal R^{(2)}_{\ket{W_N}}$, for all $k=1,\ldots,N-1$, which shows that the true maximum is taken by the $W$-state yielding Eq.~(\ref{app:R2WstMax}).
\end{proof}

\subsection{Criterion based on a linear combination of $\mathcal R^{(2)}$ and $\mathcal R^{(4)}$}\label{app:NikWit}

We want to show that the linear bound connecting the points $\ket{W_N}$ and $\ket{\phi}\ket{W_{N-1}}$ in the $(\mathcal R^{(2)},\mathcal R^{(4)})$-plane provides a bound for the detection of non-$W$-class states. To do so, it is useful to note the following relations:
\begin{align}
\mathcal R^{(4)}_{\ket{W_N} }&= \frac{83 N^3 + 216 N^2-176 N-96}{27N^3 5^N},
\end{align}
and
\begin{align}
  \mathcal R^{(2)}_{\ket{\phi}\ket{W_{N-1}}} &=\frac{1}{3} \mathcal R^{(2)}_{\ket{W_{N-1}} }, \\
  \mathcal R^{(4)}_{\ket{\phi}\ket{W_{N-1}}} &=\frac{1}{5} \mathcal R^{(4)}_{\ket{W_{N-1}} }.
\end{align}
A line passing through the $W$-state $\ket{W_N}$ and the bi-seprable state $\ket\phi\ket{W_{N-1}}$ in the $(\mathcal R^{(2)},\mathcal R^{(4)})$-plane can then be defined as follows
\begin{align}
\tilde{R}_4= m\tilde{R}_2 - b,
\end{align}
with the rescaled moments $\tilde{R}_t=(t-1)^N\mathcal R^{(t)}$, and 
\begin{align}
    m &= \frac{-54N^4+196N^3-114N^2-28 N+24}{27N^2(N-1)^2}, \label{eq:slope} \\
    b &= \frac{353N^4-1146N^3+829N^2+156 N-216}{27N^2(N-1)^2}.
\end{align}
Lastly, we need to prove that for all states in the $W$-class $\text{Conv}(\mathcal W^{(N)})$ the expression $\text{Lin}(\tilde{R}_2/3^N,\tilde{R}_4/5^N):=\tilde{R}_4 - m\tilde{R}_2 - b$ is non-positive. In order to  do so, we make use of two facts: first, $m<0$ for all $N\geq 2$. Thus, the expression $\tilde{R}_4 - m\tilde{R}_2 - b$  combines the two convex functions $\tilde{R}_2$ and $\tilde{R}_4$ with positive coefficients, yielding again a convex function. Hence, in order to prove the criterion it suffices to maximize $\tilde{R}_4 - m\tilde{R}_2 - b$ over pure states in the $W$-class, i.e., $\mathcal W^{(N)}$.
Second, the expression is still invariant under local unitary transformations, thus we can optimize over pure states in the standard form (\ref{app:Wstandardform}) instead.
For these states, the second moment is given in Eq.~(\ref{app:R2Wstandardform2}) and the fourth moment reads
\begin{align}
    \tilde{R}_4 &=1+\frac{16}3 \sum_{i<j}x_ix_j + \frac{128}3\sum_{i<j}x_i^2x_j^2 -\frac{448}9\sum_{i<j<k}x_ix_jx_k \\
                            &\phantom{=}+64\sum_{i\neq j,k \wedge j<k} x_i^2 x_j x_k+\frac{1664}9\sum_{i<j<k<l}x_ix_jx_kx_l.
\label{eq:R4Wst}
\end{align}
where we have already eliminated $x_0$ using the normalization constraint and all sums are assumed to iterate over $1,\ldots,N$. Due to the symmetry of the target function and the constraints, we assume in the following that $x_1 \leq x_2 \leq \ldots \leq x_N$.
In order to maximize the expression $\tilde{R}_4 - m\tilde{R}_2 - b$, we first show that for fixed $x_2,\ldots,x_N$, the optimum lies at the boundary of the set of allowed values for $x_1$, namely ${x}_1=\frac1N$ or ${x}_1=0$. To do so, we closer investigate the partial derivative of $\tilde{R}_4 - m\tilde{R}_2 - b$ with respect to $x_1$ which shows that it does not vanish in the region $[0, 1/N]$ and thus implies that the optimum lies at the boundary.  

In the first case, $x_1=x_2=\ldots=x_N=\frac1N$ due to the ordering. In the second case, inserting $x_1=0$ into $\tilde{R}_4 - m\tilde{R}_2 - b$ yields the same expression for the case of $N-1$ particles, except for the slope of $m(N)$, which still depends on the larger $N$. However, the optimization of the remaining variables follows exactly the same steps, yielding for $x_2$ the two candidates $x_2=0$ and $x_2=\frac{1}{N-1}$. Thus, we end up with $N+1$ candidate points for the optimum: (1) $x_1= \ldots =x_{N} = \frac1N$, (2) $x_1=0,\ x_2=\ldots=x_N=\frac1{N-1}$, \ldots, (N+1) $x_1=\ldots x_N = 0$. Inserting all the candidates into $\tilde{R}_4 - m\tilde{R}_2 - b$ yields zero for cases (1)~and (2)~as expected, as these cases correspond to choosing $ \ket{W_N}$ and $ \ket{0}\ket{W_{N-1}}$, and something negative for the remaining ones, thus proving the bound.

\end{appendix}



\end{document}